\documentclass[11pt,onecolumn,draftcls]{IEEEtran}
\newtheorem{theorem}{Theorem}
\newtheorem{proposition}{Proposition}
\newtheorem{corollary}{Corollary}

\newtheorem{lemma}{Lemma}

\newcommand{\df}{\stackrel{\mbox{\scriptsize def}}{=}}
\newcommand{\ELS}{ELS}
\newcommand{\rk}{\mathrm{rk}}

\newcommand{\dr}{d_{\mbox{\tiny R}}}
\newcommand{\Kr}{K_{\mbox{\tiny R}}}
\newcommand{\Ar}{A_{\mbox{\tiny R}}}

\newcommand{\vspan}[1]{\left< #1 \right>}

\usepackage{amsmath,amssymb,cite,subfigure}
\usepackage[dvips]{graphicx}
\usepackage{url}

\begin{document}
\title{Packing and Covering Properties of Rank Metric Codes}
\author{Maximilien Gadouleau and Zhiyuan Yan\\
Department of Electrical and Computer Engineering \\
Lehigh University, PA 18015, USA\\ E-mails: \{magc,
yan\}@lehigh.edu} \maketitle

\thispagestyle{empty}

\begin{abstract}
This paper investigates packing and covering properties of codes
with the rank metric. First, we investigate packing properties of
rank metric codes. Then, we study sphere covering properties of rank
metric codes, derive bounds on their parameters, and investigate
their asymptotic covering properties.
\end{abstract}

\section{Introduction}\label{sec:introduction}
Although the rank has long been known to be a metric implicitly and
explicitly (see, for example, \cite{hua_cms51}), the rank metric was
first considered for error control codes by Delsarte
\cite{delsarte_jct78}. The potential applications of rank metric
codes to wireless communications \cite{lusina_it03}, public-key
cryptosystems \cite{gabidulin_lncs91}, and storage equipments
\cite{gabidulin_pit0285, roth_it91} have motivated a steady stream
of works \cite{delsarte_jct78, gabidulin_pit0185, gabidulin_pit0285,
roth_it91, babu_phd95, chen_mn96, vasantha_gs99, vasantha_itw1002,
richter_isit04, kshevetskiy_isit05, gabidulin_isit05,
loidreau_wcc05, gadouleau_globecom06, gadouleau_itw06,
gadouleau_it06, loidreau_arxiv07}, described below, that focus on
their properties.

The majority \cite{delsarte_jct78, gabidulin_pit0185,
gabidulin_pit0285, roth_it91, kshevetskiy_isit05, richter_isit04,
loidreau_wcc05, gadouleau_itw06, gadouleau_it06} of previous works
focus on rank distance properties, code construction, and efficient
decoding of rank metric codes. Some previous works focus on the
packing and covering properties of rank metric codes. Both packing
and covering properties are significant for error control codes, and
packing and covering radii are basic geometric parameters of a code,
important in several respects \cite{delsarte_ic73}. For instance,
the covering radius can be viewed as a measure of performance: if
the code is used for error correction, then the covering radius is
the maximum weight of a correctable error vector
\cite{berger_book71}; if the code is used for data compression, then
the covering radius is a measure of the maximum distortion
\cite{berger_book71}. The Hamming packing and covering radii of
error control codes have been extensively studied (see, for
example,\cite{macwilliams_book77, cohen_book97}), whereas the rank
packing and covering radii have received relatively little
attention. It was shown that nontrivial perfect rank metric codes do
not exist in \cite{babu_phd95, chen_mn96, gadouleau_globecom06}. In
\cite{vasantha_gs99}, a sphere covering bound for rank metric codes
was introduced. Generalizing the concept of rank covering radius,
the multi-covering radii of codes with the rank metric were defined
in \cite{vasantha_itw1002}. Bounds on the volume of balls with rank
radii were also derived \cite{loidreau_arxiv07}.

In this paper, we investigate packing and covering properties of
rank metric codes. The main contributions of this paper are:
\begin{itemize}

\item In Section~\ref{sec:packing}, we study the packing
properties of rank metric codes.

\item In Section~\ref{sec:technical_results}, we establish further properties of elementary linear subspaces
\cite{gadouleau_it06}, and investigate properties of balls with rank
radii. In particular, we derive both upper and lower bounds on the
volume of balls with given rank radii, and our bounds are tighter
than their respective counterparts in \cite{loidreau_arxiv07}.

\item In Section~\ref{sec:covering}, we first derive both
upper and lower bounds on the minimal cardinality of a code with
given length and rank covering radius. Our new bounds are tighter
than the bounds introduced in \cite{vasantha_gs99}. We also
establish additional sphere covering properties for linear rank
metric codes, and prove that some classes of rank metric codes have
maximal covering radius. Finally, we establish the asymptotic
minimum code rate for a code with given relative covering radius.
\end{itemize}


\section{Preliminaries}\label{sec:preliminaries}
\subsection{Rank metric}\label{sec:rank_metric}
Consider an $n$-dimensional vector ${\bf x} = (x_0, x_1,\ldots,
x_{n-1}) \in \mathrm{GF}(q^m)^n$. The field $\mathrm{GF}(q^m)$ may
be viewed as an $m$-dimensional vector space over $\mathrm{GF}(q)$.
The rank weight of ${\bf x}$, denoted as $\rk({\bf x})$, is defined
to be the \emph{maximum} number of coordinates in ${\bf x}$ that are
linearly independent over $\mathrm{GF}(q)$ \cite{gabidulin_pit0185}.
Note that all ranks are with respect to $\mathrm{GF}(q)$ unless
otherwise specified in this paper. The coordinates of ${\bf x}$ thus
span a linear subspace of $\mathrm{GF}(q^m)$, denoted as
$\mathfrak{S}({\bf x})$ or $\mathfrak{S}(x_0,x_1,\ldots,x_{n-1})$,
with dimension equal to $\rk({\bf x})$. For any basis $B_m$ of
$\mathrm{GF}(q^m)$ over $\mathrm{GF}(q)$, each coordinate of ${\bf
x}$ can be expanded to an $m$-dimensional column vector over
$\mathrm{GF}(q)$ with respect to $B_m$. The rank weight of ${\bf x}$
is hence the rank of the $m\times n$ matrix over $\mathrm{GF}(q)$
obtained by expanding all the coordinates of ${\bf x}$. For all
${\bf x}, {\bf y}\in \mathrm{GF}(q^m)^n$, it is easily verified that
$\dr({\bf x},{\bf y})\df \rk({\bf x} - {\bf y})$ is a metric over
GF$(q^m)^n$ \cite{gabidulin_pit0185}, referred to as the \emph{rank
metric} henceforth. The {\em minimum rank distance} of a code
${\mathcal C}$, denoted as $d_{\mbox{\tiny R}}({\mathcal C})$, is
simply the minimum rank distance over all possible pairs of distinct
codewords. When there is no ambiguity about ${\mathcal C}$, we
denote the minimum rank distance as $\dr$.

Both the matrix form \cite{delsarte_jct78, roth_it91} and the vector
form \cite{gabidulin_pit0185} for rank metric codes have been
considered in the literature. Following \cite{gabidulin_pit0185}, in
this paper the vector form over $\mathrm{GF}(q^m)$ is used for rank
metric codes although their rank weight can be defined by their
corresponding $m \times n$ code matrices over $\mathrm{GF}(q)$
\cite{gabidulin_pit0185}. The vector form is chosen in this paper
since our results and their derivations for rank metric codes can be
readily related to their counterparts for Hamming metric codes.

\subsection{Sphere packing and sphere covering}\label{sec:covering_radius}

The sphere packing problem we consider is as follows: given a finite
field $\mathrm{GF}(q^m)$, length $n$, and radius $r$, what is the
maximum number of non-intersecting balls with radius $r$ that can be
packed into $\mathrm{GF}(q^m)^n$? The sphere packing problem is
equivalent to finding the maximum cardinality of a code over
$\mathrm{GF}(q^m)$ with length $n$ and minimum distance $d \geq
2r+1$: the spheres of radius $r$ centered at the codewords of such a
code do not intersect one another. Furthermore, when these
non-intersecting spheres centered at all codewords cover the {\em
whole} space, the code is called a perfect code.

The covering radius $\rho$ of a code $C$ with length $n$ over
$\mathrm{GF}(q^m)$ is defined to be the smallest integer $\rho$ such
that all vectors in the space $\mathrm{GF}(q^m)^n$ are within
distance $\rho$ of some codeword of $C$ \cite{cohen_book97}. It is
the maximal distance from any vector in $\mathrm{GF}(q^m)^n$  to the
code $C$. That is, $\rho = \max_{{\bf x} \in \mathrm{GF}(q^m)^n} \{
d({\bf x},C)\}$. Also, if $C \subset C'$, then the covering radius
of $C$ is no less than the minimum distance of $C'$. Finally, a code
$C$ with length $n$ and minimum distance $d$ is called a maximal
code if there does not exist any code $C'$ with the same length and
minimum distance such that $C \subset C'$. A maximal code has
covering radius $\rho \leq d-1$. The sphere covering problem for the
rank metric can be stated as follows: given an extension field
$\mathrm{GF}(q^m)$, length $n$, and radius $\rho$, we want to
determine the minimum number of balls of rank radius $\rho$ which
cover $\mathrm{GF}(q^m)^n$ entirely. The sphere covering problem is
equivalent to finding the minimum cardinality of a code over
$\mathrm{GF}(q^m)$ with length $n$ and rank covering radius $\rho$.

\section{Packing properties of rank metric
codes}\label{sec:packing}

It can be shown that the cardinality $K$ of a code $C$ over
$\mathrm{GF}(q^m)$ with length $n$ and minimum rank distance $\dr$
satisfies $K \leq \min \left\{q^{m(n-\dr+1)},q^{n(m-\dr+1)}
\right\}$. We refer to this bound as the Singleton
bound\footnote{The Singleton bound in \cite{roth_it91} has a
different form since array codes are defined over base fields.} for
codes with the rank metric, and refer to codes that attain the
Singleton bound as maximum rank distance (MRD) codes.

For any given parameter set $n$, $m$, and $\dr$, explicit
construction for linear or nonlinear MRD codes exists. For $n\leq m$
and $\dr \leq n$, generalized Gabidulin codes
\cite{kshevetskiy_isit05} can be constructed. For $n>m$ and $\dr
\leq m$, an MRD code can be constructed by transposing a generalized
Gabidulin code of length $m$ and minimum rank distance $\dr$ over
GF$(q^n)$, although this code is not necessarily linear over
GF$(q^m)$. When $n=lm$ ($l\geq 2$), linear MRD codes of length $n$
and minimum distance $\dr$ can be constructed by a cartesian product
$\mathcal{G}^l$ of an $(m,k)$ generalized Gabidulin code
$\mathcal{G}$. Although maximum distance separable codes, which
attain the Singleton bound for the Hamming metric, exist only for
limited block length over any given field, MRD codes can be
constructed for any block length $n$ and minimum rank distance $\dr$
over arbitrary fields GF$(q^m)$. This has significant impact on the
packing properties of rank metric codes as explained below.

For the Hamming metric, although nontrivial perfect codes do exist,
the optimal solution to the sphere packing problem is not known for
all the parameter sets \cite{macwilliams_book77}. In contrast, for
rank metric codes, although nontrivial perfect rank metric codes do
not exist \cite{babu_phd95, chen_mn96}, MRD codes provide an optimal
solution to the sphere packing problem for any set of parameters.
For given $n$, $m$, and $r$, let us denote the maximum cardinality
among rank metric codes over $\mathrm{GF}(q^m)$ with length $n$ and
minimum distance $\dr = 2r+1$ as $A_{\mbox{\tiny R}}(q^m,n,\dr)$.
Thus, for $\dr > \min\{n,m\}$, $A_{\mbox{\tiny R}}(q^m,n,\dr) = 1$
and for $\dr \leq \min\{n,m\}$, $A_{\mbox{\tiny R}}(q^m,n,\dr) =
\min \left\{ q^{m(n-\dr+1)}, q^{n(m-\dr+1)}\right\}$. Note that the
maximal cardinality is achieved by MRD codes for all parameter sets.
Hence, MRD codes admit the optimal solutions to the sphere packing
problem for rank metric codes.

%
%

\section{Technical results}\label{sec:technical_results}

\subsection{Further properties of elementary linear subspaces}\label{sec:lemmas_ELS}

The concept of elementary linear subspace was introduced in our
previous work \cite{gadouleau_it06}. It has similar properties to
those of a set of coordinates, and as such has served as a useful
tool in our derivation of properties of Gabidulin codes (see
\cite{gadouleau_it06}). Although our results may be derived without
the concept of ELS, we have adopted it in this paper since it
enables readers to easily relate our approach and results to their
counterparts for Hamming metric codes.

If there exists a basis set $B$ of vectors in $\mathrm{GF}(q)^n$ for
a linear subspace $\mathcal{V} \subseteq \mathrm{GF}(q^m)^n$, we say
$\mathcal{V}$ is an elementary linear subspace and $B$ is an
elementary basis of $\mathcal{V}$. We denote the set of all \ELS{}'s
of $\mathrm{GF}(q^m)^n$ with dimension $v$ as $E_v(q^m,n)$. The
properties of an ELS are summarized as follows
\cite{gadouleau_it06}. A vector has rank $\leq r$ if and only if it
belongs to some ELS with dimension $r$. For any $\mathcal{V} \in
E_v(q^m,n)$, there exists $\bar{\mathcal{V}} \in E_{n-v}(q^m,n)$
such that $\mathcal{V} \oplus \bar{\mathcal{V}} =
\mathrm{GF}(q^m)^n$, where $\oplus$ denotes the direct sum of two
subspaces. For any vector ${\bf x} \in \mathrm{GF}(q^m)^n$, we
denote the projection of ${\bf x}$ on $\mathcal{V}$ along
$\bar{\mathcal{V}}$ as ${\bf x}_\mathcal{V}$, and we remark that
${\bf x} = {\bf x}_\mathcal{V} + {\bf x}_{\bar{\mathcal{V}}}$.

In order to simplify notations, we shall occasionally denote the
vector space $\mathrm{GF}(q^m)^n$ as $F$. We denote the number of
vectors of rank $u$ ($0 \leq u \leq \min\{m,n\}$) in
$\mathrm{GF}(q^m)^n$ as $N_u(q^m,n)$. It can be shown that
$N_u(q^m,n) = {n \brack u} \alpha(m,u)$ \cite{gabidulin_pit0185},
where $\alpha(m,0) \df 1$ and $\alpha(m,u) \df
\prod_{i=0}^{u-1}(q^m-q^i)$ for $u \geq 1$. The ${n \brack u}$ term
is often referred to as a Gaussian polynomial \cite{andrews_book76},
defined as ${n \brack u} \df \alpha(n,u)/\alpha(u,u)$. Note that
$|E_u(q^m,n)| = {n \brack u}$ does not depend on $m$.

\begin{lemma}\label{lemma:unique_ELS}
Any vector ${\bf x} \in \mathrm{GF}(q^m)^n$ with rank $r$ belongs to
a unique \ELS{} $\mathcal{V} \in E_r(q^m,n)$.
\end{lemma}

\begin{proof}
The existence of $\mathcal{V} \in E_r(q^m,n)$ has been proved in
\cite{gadouleau_it06}. Thus we only prove the uniqueness of
$\mathcal{V}$, with elementary basis $\{ {\bf v}_i \}_{i=0}^{r-1}$,
where ${\bf v}_i \in \mathrm{GF}(q)^n$ for all $i$. Suppose ${\bf
x}$ also belongs to $\mathcal{W}$, where $\mathcal{W} \in
E_r(q^m,n)$ has an elementary basis $\{ {\bf w}_j \}_{j=0}^{r-1}$,
where ${\bf w}_j \in \mathrm{GF}(q)^n$ for all $j$. Therefore, ${\bf
x} = \sum_{i=0}^{r-1} a_i {\bf v}_i = \sum_{j=0}^{r-1} b_j {\bf
w}_j$, where $a_i,b_j \in \mathrm{GF}(q^m)$ for $0 \leq i,j \leq
r-1$. By definition, we have $\mathfrak{S}({\bf x}) =
\mathfrak{S}(a_0,\ldots,a_{r-1}) =
\mathfrak{S}(b_0,\ldots,b_{r-1})$, therefore $b_j$'s can be
expressed as linear combinations of $a_i$'s, i.e., $b_j =
\sum_{i=0}^{r-1} c_{j,i} a_i$ where $c_{j,i} \in \mathrm{GF}(q)$.
Hence ${\bf x} = \sum_{i=0}^{r-1} a_i {\bf u}_i$, where ${\bf u}_i =
\sum_{j=0}^{r-1} c_{j,i} {\bf w}_j$ for $0 \leq i \leq r-1$ form an
elementary basis of $\mathcal{W}$. Considering the matrix obtained
by expanding the coordinates of ${\bf x}$ with respect to the basis
$\{a_i\}_{i=0}^{m-1}$, we obtain ${\bf v}_i = {\bf u}_i$, and hence
$\mathcal{V} = \mathcal{W}$.
\end{proof}

Lemma~\ref{lemma:unique_ELS} shows that an ELS is analogous to a
subset of coordinates since a vector ${\bf x}$ with Hamming weight
$r$ belongs to a unique subset of $r$ coordinates, often referred to
as the support of ${\bf x}$.

In \cite{gadouleau_it06}, it was shown that an \ELS{} always has a
complementary elementary linear subspace. The following lemma
enumerates such complementary ELS's.
\begin{lemma}\label{lemma:num_B}
Suppose $\mathcal{V} \in E_v(q^m,n)$ and $\mathcal{A} \subseteq
\mathcal{V}$ is an \ELS{} with dimension $a$, then there are
$q^{a(v-a)}$ \ELS{}'s $\mathcal{B}$ such that $\mathcal{A} \oplus
\mathcal{B} = \mathcal{V}$. Furthermore, there are $q^{a(v-a)} {v
\brack a}$ such ordered pairs $(\mathcal{A},\mathcal{B})$.
\end{lemma}

\begin{proof}
First, remark that $\mathrm{dim}(\mathcal{B}) = v-a$. The total
number of sets of $v-a$ linearly independent vectors over GF$(q)$ in
$\mathcal{V} \backslash \mathcal{A}$ is given by $N = (q^v -
q^a)(q^v-q^{a+1}) \cdots (q^v-q^{v-1}) = q^{a(v-a)}\alpha(v-a,v-a)$.
Note that each set of linearly independent vectors over GF$(q)$
constitutes an elementary basis set. Thus, the number of possible
$\mathcal{B}$ is given by $N$ divided by $\alpha(v-a,v-a)$, the
number of elementary basis sets for each $\mathcal{B}$. Therefore,
once $\mathcal{A}$ is fixed, there are $q^{a(v-a)}$ choices for
$\mathcal{B}$. Since the number of $a$-dimensional subspaces
$\mathcal{A}$ in $\mathcal{V}$ is ${v \brack a}$, the total number
of ordered pairs $(\mathcal{A},\mathcal{B})$ is hence $q^{a(v-a)} {v
\brack a}$.
\end{proof}

Puncturing a vector with full Hamming weight results in another
vector with full Hamming weight. Lemma~\ref{lemma:restriction_u}
below shows that the situation for vectors with full rank is
similar.

\begin{lemma}\label{lemma:restriction_u}
Suppose $\mathcal{V} \in E_v(q^m,n)$ and ${\bf u} \in \mathcal{V}$
has rank $v$, then $\rk({\bf u}_\mathcal{A}) = a$ and $\rk({\bf
u}_\mathcal{B}) = v-a$ for any $\mathcal{A} \in E_a(q^m,n)$ and
$\mathcal{B} \in E_{v-a}(q^m,n)$ such that $\mathcal{A} \oplus
\mathcal{B} = \mathcal{V}$.
\end{lemma}

\begin{proof}
First, ${\bf u}_\mathcal{A} \in \mathcal{A}$ and hence $\rk({\bf
u}_\mathcal{A}) \leq a$ by \cite[Proposition~2]{gadouleau_it06};
similarly, $\rk({\bf u}_\mathcal{B}) \leq v-a$. Now suppose
$\rk({\bf u}_\mathcal{A}) < a$ or $\rk({\bf u}_\mathcal{B}) < v-a$,
then $v = \rk({\bf u}) \leq \rk({\bf u}_\mathcal{A}) + \rk({\bf
u}_\mathcal{B}) < a+v-a = v$.
\end{proof}

It was shown in \cite{gadouleau_it06} that the projection ${\bf
u}_\mathcal{A}$ of a vector ${\bf u}$ on an \ELS{} $\mathcal{A}$
depends on both $\mathcal{A}$ and its complement $\mathcal{B}$. The
following lemma further clarifies the relation: changing
$\mathcal{B}$ always modifies ${\bf u}_\mathcal{A}$, provided that
${\bf u}$ has full rank.

\begin{lemma}\label{lemma:restriction}
Suppose $\mathcal{V} \in E_v(q^m,n)$ and ${\bf u} \in \mathcal{V}$
has rank $v$.  For any $\mathcal{A} \in E_a(q^m,n)$ and $\mathcal{B}
\in E_{v-a}(q^m,n)$ such that $\mathcal{A} \oplus \mathcal{B} =
\mathcal{V}$, define the functions $f_{\bf
u}(\mathcal{A},\mathcal{B}) = {\bf u}_\mathcal{A}$ and $g_{\bf
u}(\mathcal{A},\mathcal{B}) ={\bf u}_\mathcal{B}$. Then both $f_{\bf
u}$ and $g_{\bf u}$ are injective.
\end{lemma}

\begin{proof}
Consider another pair $(\mathcal{A}',\mathcal{B}')$ with dimensions
$a$ and $v-a$ respectively. Suppose $\mathcal{A}' \neq \mathcal{A}$,
then ${\bf u}_{\mathcal{A}'} \neq {\bf u}_\mathcal{A}$. Otherwise
${\bf u}_\mathcal{A}$ belongs to two distinct \ELS{}'s with
dimension $a$, which contradicts Lemma~\ref{lemma:unique_ELS}. Hence
${\bf u}_{\mathcal{A}'} \neq {\bf u}_\mathcal{A}$ and ${\bf
u}_{\mathcal{B}'} = {\bf u} - {\bf u}_{\mathcal{A}'} \neq {\bf u} -
{\bf u}_{\mathcal{A}} = {\bf u}_\mathcal{B}$. The argument is
similar if $\mathcal{B}' \neq \mathcal{B}$.
\end{proof}

\subsection{Properties of balls with rank radii}\label{sec:balls}

We refer to all vectors in $\mathrm{GF}(q^m)^n$ within rank distance
$r$ of ${\bf x} \in \mathrm{GF}(q^m)^n$ as a ball of rank radius $r$
centered at ${\bf x}$, and denote it as $B_r({\bf x})$. Its volume,
which does not depend on ${\bf x}$, is denoted as $V_r(q^m,n) =
\sum_{u=0}^r N_u(q^m,n)$. When there is no ambiguity about the
vector space, we denote $V_r(q^m,n)$ as $v(r)$.

\begin{lemma}\label{lemma:lower_bound_Vt}
For $0 \leq r \leq \min\{n,m\}$, $q^{r(m+n-r)} \leq V_r(q^m,n) <
K_q^{-1} q^{r(m+n-r)}$, where $K_q \df \prod_{j=1}^\infty
(1-q^{-j})$ \cite{gadouleau_it06}.
\end{lemma}

\begin{proof}
The upper bound was derived in \cite[Lemma 13]{gadouleau_it06}, and
it suffices to prove the lower bound. Without loss of generality, we
assume that the center of the ball is ${\bf 0}$. We now prove the
lower bound by constructing $q^{r(m+n-r)}$ vectors ${\bf z} \in
\mathrm{GF}(q^m)^n$ of rank at most $r$. Let ${\bf x} \in
\mathrm{GF}(q^m)^r$ let a subspace $\mathfrak{T}$ of
$\mathrm{GF}(q^m)$ such that $\mathrm{dim}(\mathfrak{T}) = r$ and
$\mathfrak{S}({\bf x}) \subseteq \mathfrak{T}$. We consider the
vectors ${\bf y} \in \mathrm{GF}(q^m)^{n-r}$ such that
$\mathfrak{S}({\bf y}) \subseteq \mathfrak{T}$. There are $q^{mr}$
choices for ${\bf x}$ and, for a given ${\bf x}$, $q^{r(n-r)}$
choices for ${\bf y}$. Thus the total number of vectors ${\bf z} =
({\bf x},{\bf y}) \in \mathrm{GF}(q^m)^n$ is $q^{r(m+n-r)}$, and
since $\mathfrak{S}({\bf z}) \subseteq \mathfrak{T}$, we have
$\rk({\bf z}) \leq r$.
\end{proof}

We remark that both bounds in Lemma~\ref{lemma:lower_bound_Vt} are
tighter than their respective counterparts in
\cite[Proposition~1]{loidreau_arxiv07}. More importantly, the two
bounds in Lemma~\ref{lemma:lower_bound_Vt} differ only by a factor
of $K_q$, and thus they not only provide a good approximation of
$V_r(q^m,n)$, but also accurately describe the asymptotic behavior
of $V_r(q^m,n)$.

The diameter of a set is defined to be the maximum distance between
any pair of elements in the set \cite[p. 172]{macwilliams_book77}.
For a binary vector space $\mathrm{GF}(2)^n$ and a given diameter
$2r<n$, Kleitman \cite{kleitman_jct66} proved that balls with
Hamming radius $r$ maximize the cardinality of a set with a given
diameter. However, when the underlying field for the vector space is
not $\mathrm{GF}(2)$, the result is not necessarily valid \cite[p.
40]{cohen_book97}. We show below that balls with rank radii do not
maximize the cardinality of a set with a given diameter.

\begin{proposition}\label{prop:counter_Kleitman}
For $2 \leq 2r \leq n \leq m$, any $S \in E_{2r}(q^m,n)$ has
diameter $2r$ and cardinality $|S| > V_r(q^m,n)$.
\end{proposition}

\begin{proof}
Any $S \in E_{2r}(q^m,n)$ has diameter $2r$ and cardinality
$q^{2mr}$. For $r=1$, we have $V_1(q^m,n) = 1 +
\frac{(q^n-1)(q^m-1)}{(q-1)} < q^{2m}$. For $r \geq 2$, we have
$V_r(q^m,n) < K_q^{-1} q^{r(n+m)-r^2}$ by
Lemma~\ref{lemma:lower_bound_Vt}. Since $r^2 > 2 > -\log_q K_q$, we
obtain $V_r(q^m,n) < q^{r(n+m)} \leq |S|$.
\end{proof}

The intersection of balls with Hamming radii has been studied in
\cite[Chapter 2]{cohen_book97}, and below we investigate the
intersection of balls with rank radii.

\begin{lemma}\label{lemma:inter_2_balls}
If $0 \leq r,s \leq n$ and ${\bf c}_1, {\bf c}_2 \in
\mathrm{GF}(q^m)^n$, then $|B_r({\bf c}_1) \cap B_s({\bf c}_2)|$
depends on ${\bf c}_1$ and ${\bf c}_2$ only through $\dr({\bf
c}_1,{\bf c}_2)$.
\end{lemma}

\begin{proof}
This follows from the fact that matrices in $\mathrm{GF}(q)^{m
\times n}$ together with the rank metric form an association scheme
\cite{delsarte_jct78, delsarte_it98}.
\end{proof}

\begin{proposition}\label{prop:inter_2_balls}
If $0 \leq r,s \leq n$, ${\bf c}_1, {\bf c}_2, {\bf c}'_1, {\bf
c}'_2 \in \mathrm{GF}(q^m)^n$ and $\dr({\bf c}_1,{\bf c}_2)
> \dr({\bf c}'_1,{\bf c}'_2)$, then $|B_r({\bf c}_1) \cap B_s({\bf c}_2)| \leq
|B_r({\bf c}'_1) \cap B_s({\bf c}'_2)|$.
\end{proposition}

\begin{proof}
It suffices to prove the claim when $\dr({\bf c}_1,{\bf c}_2) =
\dr({\bf c}'_1,{\bf c}'_2) + 1 = e + 1$. By
Lemma~\ref{lemma:inter_2_balls}, we can assume without loss of
generality that ${\bf c}_1 = {\bf c}'_1 = {\bf 0}$, ${\bf c}'_2 =
(0, c_1,\ldots,c_e,0,\ldots,0)$ and ${\bf c}_2 = (c_0,
c_1,\ldots,c_e,0,\ldots,0)$, where $c_0,\ldots,c_e \in
\mathrm{GF}(q^m)$ are linearly independent.

We will show that an injective mapping $\phi$ from $B_r({\bf c}_1)
\cap B_s({\bf c}_2)$ to $B_r({\bf c}'_1) \cap B_s({\bf c}'_2)$ can
be constructed. We consider vectors ${\bf z} =
(z_0,z_1,\ldots,z_{n-1}) \in B_r({\bf c}_1) \cap B_s({\bf c}_2)$. We
thus have $\rk({\bf z}) \leq r$ and $\rk({\bf u}) \leq s$, where
${\bf u} = (u_0,u_1,\ldots,u_{n-1}) = {\bf z} - {\bf c}_2 =
(z_0-c_0,z_1-c_1,\ldots,z_{n-1})$. We also define $\bar{\bf z} =
(z_1,\ldots,z_{n-1})$ and $\bar{\bf u} = (u_1,\ldots,u_{n-1})$. We
consider three cases for the mapping $\phi$, depending on $\bar{\bf
z}$ and $\bar{\bf u}$.
\begin{itemize}
\item Case I: $\rk(\bar{\bf u}) \leq s-1$. In this case, $\phi({\bf z}) \df {\bf z}$.
We remark that $\rk({\bf z} - {\bf c}'_2) \leq \rk(\bar{\bf u}) + 1
\leq s$ and hence $\phi({\bf z}) \in B_r({\bf c}'_1) \cap B_s({\bf
c}'_2)$.

\item Case II: $\rk(\bar{\bf u}) = s$ and $\rk(\bar{\bf z}) \leq r-1$.
In this case, $\phi({\bf z}) \df (z_0-c_0,z_1,\ldots,z_{n-1})$. We
have $\rk(\phi({\bf z})) \leq \rk(\bar{\bf z}) +1 \leq r$ and
$\rk\left(\phi({\bf z}) - {\bf c}'_2\right) = \rk\left({\bf z}- {\bf
c}_2\right)\leq s$, and hence $\phi({\bf z}) \in B_r({\bf c}'_1)
\cap B_s({\bf c}'_2)$.

\item Case III: $\rk(\bar{\bf u}) = s$ and $\rk(\bar{\bf z}) = r$.
Since $\rk({\bf u}) = s$, we have $z_0 - c_0 \in
\mathfrak{S}(\bar{\bf u})$. Similarly, since $\rk({\bf z}) = r$, we
have $z_0 \in \mathfrak{S}(\bar{\bf z})$. Denote
$\dim(\mathfrak{S}(\bar{\bf u},\bar{\bf z}))$ as $d$ ($d\geq s$).
For $d
> s$, let $\alpha_0,\ldots,\alpha_{d-1}$ be a basis of
$\mathfrak{S}(\bar{\bf u},\bar{\bf z})$ such that
$\alpha_0,\ldots,\alpha_{s-1} \in \mathfrak{S}(\bar{\bf u})$ and
$\alpha_s,\ldots,\alpha_{d-1} \in \mathfrak{S}(\bar{\bf z})$. This
basis is fixed for all vectors ${\bf z}$ having the same $\bar{\bf
z}$, i.e., it is fixed for all values of $z_0$. Note that $c_0 \in
\mathfrak{S}(\bar{\bf u},\bar{\bf z})$, and may therefore be
uniquely expressed as $c_0 = c_u + c_z$, where $c_u \in
\mathfrak{S}(\alpha_0,\ldots,\alpha_{s-1}) = \mathfrak{S}(\bar{\bf
u})$ and $c_z \in
\mathfrak{S}(\alpha_s,\ldots,\alpha_{d-1})\subseteq
\mathfrak{S}(\bar{\bf z})$. If $d=s$, then $c_z=0\in
\mathfrak{S}(\bar{\bf z})$. In this case, $\phi({\bf z}) \df (z_0 -
c_z,z_1,\ldots,z_{n-1})$. Remark that $z_0-c_z \in
\mathfrak{S}(\bar{\bf z})$ and hence $\rk(\phi({\bf z})) = r$. Also,
$z_0-c_z = z_0 - c_0 + c_u \in \mathfrak{S}(\bar{\bf u})$ and hence
$\rk(\phi({\bf z}) - {\bf c}'_2) = s$. Therefore $\phi({\bf z}) \in
B_r({\bf c}'_1) \cap B_s({\bf c}'_2)$.
\end{itemize}

It can be easily verified that $\phi$ is injective, hence $|B_r({\bf
c}_1) \cap B_s({\bf c}_2)| \leq |B_r({\bf c}'_1) \cap B_s({\bf
c}'_2)|$.
\end{proof}

\begin{corollary}\label{cor:union_2_balls}
If $0 \leq r,s \leq n$, ${\bf c}_1, {\bf c}_2, {\bf c}'_1, {\bf
c}'_2 \in \mathrm{GF}(q^m)^n$ and $\dr({\bf c}_1,{\bf c}_2) \geq
\dr({\bf c}'_1,{\bf c}'_2)$, then $|B_r({\bf c}_1) \cup B_s({\bf
c}_2)| \geq |B_r({\bf c}'_1) \cup B_s({\bf c}'_2)|$.
\end{corollary}

\begin{proof}
The result follows from $|B_r({\bf c}_1) \cup B_s({\bf c}_2)| = v(r)
+ v(s) - |B_r({\bf c}_1) \cap B_s({\bf c}_2)|$.
\end{proof}

We now quantify the volume of the intersection of two balls with
rank radii for some special cases, which will be used in
Section~\ref{sec:lower_bounds}.

\begin{proposition}\label{prop:B_r}
If ${\bf c}_1, {\bf c}_2 \in \mathrm{GF}(q^m)^n$ and $\dr({\bf c}_1,
{\bf c}_2) = r$, then $|B_r({\bf c}_1) \cap B_1({\bf c}_2)| = 1 +
(q^m-q^r){r \brack 1} + (q^r-1){n \brack 1}$.
\end{proposition}

\begin{proof}
By Lemma~\ref{lemma:unique_ELS}, the vector ${\bf c}_1$ belongs to a
unique \ELS{} $\mathcal{V} \in E_r(q^m,n)$. First of all, it is easy
to check that ${\bf y} = {\bf 0} \in B_r({\bf c}_1) \cap B_1({\bf
0})$. We now consider a nonzero vector ${\bf y} \in B_1({\bf 0})$
with rank $1$. We have $\dr({\bf y}, {\bf c}_1) = r+1$ if and only
if ${\bf y} \notin \mathcal{V}$ and $\mathfrak{S}({\bf y})
\nsubseteq \mathfrak{S}({\bf c}_1)$. There are $\frac{(q^n-q^r)(q^m
- q^r)}{q-1}$ such vectors. Thus, $|B_r({\bf c}_1) \cap B_1({\bf
c}_2)| = 1 + N_1(q^m,n) - \frac{(q^n-q^r)(q^m-q^r)}{q-1} = 1 +
(q^m-q^r){r \brack 1} + (q^r-1){n \brack 1}$.
\end{proof}

\begin{proposition}\label{prop:v-a_problem}
If ${\bf c}_1, {\bf c}_2 \in \mathrm{GF}(q^m)^n$ and $\dr({\bf c}_1,
{\bf c}_2) = r$, then $|B_s({\bf c}_1) \cap B_{r-s}({\bf c}_2)| =
q^{s(r-s)} {r \brack s}$ for $0 \leq s \leq r$.
\end{proposition}

\begin{proof}
By Lemma~\ref{lemma:inter_2_balls}, we can assume that ${\bf c}_1 =
{\bf 0}$, and hence $\rk({\bf c}_2) = r$. By
Lemma~\ref{lemma:unique_ELS}, ${\bf c}_2$ belongs to a unique \ELS{}
$\mathcal{V} \in E_r(q^m,n)$. We first prove that all vectors ${\bf
y} \in B_s({\bf 0}) \cap B_{r-s}({\bf c}_2)$ are in $\mathcal{V}$.
Let ${\bf y} = {\bf y}_\mathcal{V} + {\bf y}_\mathcal{W}$, where
$\mathcal{W} \in E_{n-r}(q^m,n)$ such that $\mathcal{V} \oplus
\mathcal{W} = \mathrm{GF}(q^m)^n$. We have ${\bf y}_\mathcal{V} +
({\bf c}_2-{\bf y})_\mathcal{V} = {\bf c}_2$, with $\rk({\bf
y}_\mathcal{V}) \leq \rk({\bf y}) \leq s$ and $\rk(({\bf c}_2 - {\bf
y})_\mathcal{V}) \leq \rk({\bf c}_2 - {\bf y}) \leq r-s$. Therefore,
$\rk({\bf y}_\mathcal{V}) = \rk({\bf y}) = s$, $\rk(({\bf c}_2 -
{\bf y})_\mathcal{V}) = \rk({\bf c}_2 - {\bf y})= r-s$, and
$\mathfrak{S}({\bf y}_\mathcal{V}) \cap \mathfrak{S}(({\bf c}_2 -
{\bf y})_\mathcal{V}) = \{0 \}$. Since $\rk({\bf y}_\mathcal{V}) =
\rk({\bf y})$, we have $\mathfrak{S}({\bf y}_\mathcal{W}) \subseteq
\mathfrak{S}({\bf y}_\mathcal{V})$; and similarly
$\mathfrak{S}(({\bf c}_2 - {\bf y})_\mathcal{W}) \subseteq
\mathfrak{S}(({\bf c}_2 -{\bf y})_\mathcal{V})$. Altogether, we
obtain $\mathfrak{S}({\bf y}_\mathcal{W}) \cap \mathfrak{S}(({\bf
c}_2 - {\bf y})_\mathcal{W}) = \{0\}$. However, ${\bf y}_\mathcal{W}
+ ({\bf c}_2 - {\bf y})_\mathcal{W} = {\bf 0}$, and hence ${\bf
y}_\mathcal{W} = ({\bf c}_2 - {\bf y})_\mathcal{W} = {\bf 0}$.
Therefore, ${\bf y} \in \mathcal{V}$.

We now prove that ${\bf y}$ is necessarily the projection of ${\bf
c}_2$ onto some \ELS{} $\mathcal{A}$ of $\mathcal{V}$. If ${\bf y}
\in \mathcal{V}$ satisfies $\rk({\bf y}) = s$ and $\rk({\bf
c}_2-{\bf y}) = r-s$, then ${\bf y}$ belongs to some \ELS{}
$\mathcal{A}$ and ${\bf c}_2 - {\bf y} \in \mathcal{B}$ such that
$\mathcal{A} \oplus \mathcal{B} = \mathcal{V}$. We hence have ${\bf
y} = {\bf c}_{2,\mathcal{A}}$ and ${\bf c}_2 - {\bf y} = {\bf
c}_{2,\mathcal{B}}$.

On the other hand, for any $\mathcal{A} \in E_s(q^m,n)$ and
$\mathcal{B} \in E_{r-s}(q^m,n)$ such that $\mathcal{A} \oplus
\mathcal{B} = \mathcal{V}$, ${\bf c}_{2,\mathcal{A}}$ is a vector of
rank $s$ with distance $r-s$ from ${\bf c}_2$ by
Lemma~\ref{lemma:restriction_u}. By Lemma~\ref{lemma:restriction},
all the ${\bf c}_{2,\mathcal{A}}$ vectors are distinct. There are
thus as many vectors ${\bf y}$ as ordered pairs
$(\mathcal{A},\mathcal{B})$. By Lemma~\ref{lemma:num_B}, there are
$q^{s(r-s)}{r \brack s}$ such pairs, and hence $q^{s(r-s)}{r \brack
s}$ vectors ${\bf y}$.
\end{proof}

The problem of the intersection of three balls with rank radii is
more complicated since the volume of the intersection of three balls
with rank radii is not completely determined by the pairwise
distances between the centers. We give a simple example to
illustrate this point: consider $\mathrm{GF}(2^2)^3$ and the vectors
${\bf c}_1 = {\bf c}'_1 = (0,0,0)$, ${\bf c}_2 = {\bf c}'_2 = (1,
\alpha,0)$, ${\bf c}_3 = (\alpha,0,1)$, and ${\bf c}'_3 =
(\alpha,\alpha+1,0)$, where $\alpha$ is a primitive element of the
field. It can be verified that $\dr({\bf c}_1,{\bf c}_2)=\dr({\bf
c}_2,{\bf c}_3)=\dr({\bf c}_3,{\bf c}_1)=2$ and $\dr({\bf c}'_1,{\bf
c}'_2)=\dr({\bf c}'_2,{\bf c}'_3)=\dr({\bf c}'_3,{\bf c}'_1)=2$.
However, $B_1({\bf c}_1) \cap B_1({\bf c}_2) \cap B_1({\bf c}_3) =
\{(\alpha+1,0,0)\}$, whereas $B_1({\bf c}'_1) \cap B_1({\bf c}'_2)
\cap B_1({\bf c}'_3) = \{(1,0,0), (0,\alpha+1,0),
(\alpha,\alpha,0)\}$. We remark that this is similar to the problem
of the intersection of three balls with Hamming radii discussed in
\cite[p. 58]{cohen_book97}, provided that the underlying field is
not $\mathrm{GF}(2)$.

\section{Covering properties of rank metric
codes}\label{sec:covering}

\subsection{The sphere covering problem}\label{sec:covering_intro}

We denote the minimum cardinality of a code of length $n$ and rank
covering radius $\rho$ as $\Kr(q^m,n,\rho)$. We remark that if $C$
is a code over $\mathrm{GF}(q^m)$ with length $n$ and covering
radius $\rho$, then its transpose code $C^T$ is a code over
$\mathrm{GF}(q^n)$ with length $m$ and the same covering radius.
Therefore, $K_{\mbox{\tiny R}}(q^m,n,\rho) = K_{\mbox{\tiny
R}}(q^n,m,\rho)$, and without loss of generality we shall assume $n
\leq m$ henceforth in this section. Also note that $K_{\mbox{\tiny
R}}(q^m,n,0) = q^{mn}$ and $K_{\mbox{\tiny R}}(q^m,n,n) = 1$ for all
$m$ and $n$. Hence we assume $0< \rho <n$ throughout this section.
Two bounds on $K_{\mbox{\tiny R}}(q^m,n,\rho)$ can be easily
derived.

\begin{proposition}\label{prop:obvious_bounds_K}
For $0 < \rho < n \leq m$, $\frac{q^{mn}}{v(\rho)} < \Kr(q^m,n,\rho)
\leq q^{m(n-\rho)}$.
\end{proposition}

\begin{proof}
The lower bound is a straightforward generalization of the bound
given in \cite{vasantha_gs99}. Note that the only codes with
cardinality $\frac{q^{mn}}{v(\rho)}$ are perfect codes. However,
there are no nontrivial perfect codes for the rank metric
\cite{babu_phd95}. Therefore, $K_{\mbox{\tiny R}} (q^m,n,\rho) >
\frac{q^{mn}} {v(\rho)}$. The upper bound follows from $\rho \leq
n-k$ for any $(n,k)$ linear code (see \cite{cohen_book97} for a
proof in the Hamming metric), and hence any linear code with
covering radius $\rho$ has cardinality $\leq q^{m(n-\rho)}$.
\end{proof}

We refer to the lower bound in
Proposition~\ref{prop:obvious_bounds_K} as the sphere covering
bound.

For a code over $\mathrm{GF}(q^m)$ with length $n$ and covering
radius $0 < \rho < n$, we have $K_{\mbox{\tiny R}}(q^m,n,\rho) \leq
K_{\mbox{{\tiny H}}}(q^m,n,\rho)$, where $K_{\mbox{{\tiny
H}}}(q^m,n,\rho)$ is the minimum cardinality of a (linear or
nonlinear) code over $\mathrm{GF}(q^m)$ with length $n$ and Hamming
covering radius $\rho$. This holds because any code with Hamming
covering radius $\rho$ has rank covering radius $\leq \rho$. Since
$K_{\mbox{{\tiny H}}}(q^m,n,\rho) \leq q^{m(n-\rho)}$
\cite{cohen_book97}, this provides a tighter bound than the one
given in Proposition~\ref{prop:obvious_bounds_K}.

\begin{proposition}\label{prop:K>=3}
For $0 < \rho < n \leq m$, $\Kr(q^m,n,\rho) \geq 3$.
\end{proposition}

\begin{proof}
Suppose there exists a code $C$ of cardinality $2$ and length $n$
over $\mathrm{GF}(q^m)$ with covering radius $\rho < n$. Without
loss of generality, we assume $C = \left\{ {\bf 0}, {\bf c}
\right\}$. Since $|B_\rho({\bf 0}) \cup B_\rho({\bf c})|$ is a
non-decreasing function of $\rk({\bf c})$ by
Corollary~\ref{cor:union_2_balls}, we assume $\rk({\bf c}) = n$. The
code $\mathcal{G} = \vspan{{\bf c}}$ is hence an $(n,1,n)$ linear
MRD code over $\mathrm{GF}(q^m)$. Therefore, any codeword in
$\mathcal{G} \backslash C$ is at distance $n$ from $C$. Thus $\rho =
n$, which contradicts our assumption.
\end{proof}

\begin{lemma}\label{lemma:K_n_n'}
$\Kr(q^m, n+n', \rho + \rho') \leq \Kr(q^m,n,\rho) \Kr(q^m, n',
\rho')$ for all $m>0$ and nonnegative $n$, $n'$, $\rho$, and
$\rho'$. In particular, $\Kr(q^m,n+1,\rho+1) \leq \Kr(q^m,n,\rho)$
and $\Kr(q^m,n+1,\rho) \leq q^m \Kr(q^m,n,\rho)$.
\end{lemma}

\begin{proof}
For all ${\bf x}, {\bf y} \in \mathrm{GF}(q^m)^n$ and ${\bf x}',
{\bf y}' \in \mathrm{GF}(q^m)^{n'}$, we have $\dr(({\bf x}, {\bf
x}'), ({\bf y},{\bf y}')) \leq \dr({\bf x}, {\bf y}) + \dr({\bf y},
{\bf y}')$. Therefore, for any $C \in \mathrm{GF}(q^m)^n$, $C' \in
\mathrm{GF}(q^m)^{n'}$, we have $\rho(C \oplus C') \leq \rho(C) +
\rho(C')$ and the first claim follows. In particular, $(n',\rho') =
(1,1)$ and $(n',\rho') = (1,0)$ yield the other two claims
respectively.
\end{proof}

\subsection{Lower bounds for the sphere covering
problem}\label{sec:lower_bounds}

We now derive two nontrivial lower bounds on $K_{\mbox{\tiny
R}}(q^m,n,\rho)$. For $0 \leq d \leq n$, we denote the volume of the
intersection of two balls in $\mathrm{GF}(q^m)^n$ with rank radii
$\rho$ and a distance $d$ between their respective centers $d$ as
$I(q^m,n,\rho,d)$. When there is no ambiguity about the vector space
considered, we simply denote it as $I(\rho,d)$. $I(\rho,d)$ is well
defined by Lemma~\ref{lemma:inter_2_balls}, and obviously
$I(\rho,d)=0$ when $d>2\rho$.

\begin{proposition}\label{prop:bound_cohen_generalised}
For $0 < \rho < n \leq m$ and $0 \leq l \leq \lfloor \log_{q^m}
\Kr(q^m,n,\rho) \rfloor$,
\begin{equation}\label{eq:bound_cohen_generalised}
    \Kr(q^m,n,\rho) \geq \frac{q^{mn} - q^{lm}I(\rho,n-l) +
    \sum_{a=\max\{1,n-2\rho+1\}}^l (q^{am} - q^{(a-1)m}) I(\rho,n-a+1)}
    {v(\rho) - I(\rho,n-l)}.
\end{equation}
\end{proposition}

\begin{proof}
Let us denote $\lfloor \log_{q^m} \Kr(q^m,n,\rho) \rfloor$ as
$\lambda$ for convenience. Let $C = \{ {\bf c}_i \}_{i=0}^{K-1}$ be
a code of length $n$ and covering radius $\rho$ over
$\mathrm{GF}(q^m)$. Define $C_j \df \{ {\bf c}_i \}_{i=0}^j$ for $0
\leq j \leq K-1$. For $1 \leq a \leq \lambda$ and $q^{m(a-1)} \leq j
< q^{ma}$, we have $\dr({\bf c}_j,C_{j-1}) \leq n-a+1$ by the
Singleton bound. The codeword ${\bf c}_j$ hence covers at most
$v(\rho) - I(\rho, n-a+1)$ vectors that are not previously covered
by $C_{j-1}$. For $1 \leq l \leq \lambda$, the number of vectors
covered by $C$ thus satisfies
\begin{equation}\label{eq:sum_I}
    q^{mn} \leq v(\rho) + \sum_{a=1}^l (q^{am} - q^{(a-1)m})[v(\rho) -
    I(\rho,n-a+1)] + (K-q^{lm})[v(\rho) - I(\rho,n-l)].
\end{equation}
Since $I(\rho,n-a+1) = 0$ for $a \leq n-2\rho$, (\ref{eq:sum_I})
reduces to~(\ref{eq:bound_cohen_generalised}).
\end{proof}

Note that the RHS of~(\ref{eq:bound_cohen_generalised}) is a
non-decreasing function of $l$, thus the bound is tightest when $l =
\lfloor \log_{q^m} \Kr(q^m,n,\rho) \rfloor$. We obtain a lower bound
by using the largest $l$ such that the RHS
of~(\ref{eq:bound_cohen_generalised}) is less than $q^{(l+1)m}$.

\begin{corollary}\label{cor:bound_K_I}
For $0 < \rho < n \leq m$, $\Kr(q^m,n,\rho) \geq \frac{q^{mn} -
I(\rho,n)}{v(\rho) - I(\rho,n)}$.
\end{corollary}

\begin{proof}
This is a special case of
Proposition~\ref{prop:bound_cohen_generalised} for $l=0$.
\end{proof}

\begin{corollary}\label{cor:bound_cohen}
For all $m$, $n$, and $0 < \rho \leq \lfloor n/2 \rfloor$,
$K_{\mbox{\tiny R}}(q^m,n,\rho) \geq
\frac{q^{mn}-q^{m(n-2\rho)+\rho^2} {2\rho \brack \rho}}
{v(\rho)-q^{\rho^2}{2\rho \brack \rho}}$.
\end{corollary}

\begin{proof}
Since the balls of rank radius $\rho$ around the codewords of a code
with minimum rank distance $2\rho+1$ do not intersect, we have
$\Kr(q^m,n,\rho) \geq \Ar(q^m,n,2\rho+1) = q^{m(n-2\rho)}$, and
hence $\log_{q^m} \Kr(q^m,n,\rho) \geq n-2\rho$. Use
Proposition~\ref{prop:bound_cohen_generalised} for $l = n-2\rho \geq
0$, and $I(\rho,2\rho) = q^{\rho^2}{2\rho \brack \rho}$ by
Proposition~\ref{prop:v-a_problem}.
\end{proof}

The bound in Corollary~\ref{cor:bound_cohen} can be viewed as the
counterpart in the rank metric of the bound in \cite[Theorem
1]{cohen_it86}.

Van Wee \cite{vanwee_it88, vanwee_jct91} derived several bounds on
codes with Hamming covering radii based on the excess of a code,
which is determined by the number of codewords covering the same
vectors. Although the concepts in \cite{vanwee_it88, vanwee_jct91}
were developed for the Hamming metric, they are in fact independent
of the underlying metric and thus are applicable to the rank metric
as well. For all $V \subseteq \mathrm{GF}(q^m)^n$ and a code $C$
with covering radius $\rho$, the excess on $V$ by $C$ is defined to
be $E_C(V) \df \sum_{{\bf c} \in C}|B_\rho({\bf c}) \cap V| - |V|$.
If $\{W_i\}$ is a family of disjoint subsets of
$\mathrm{GF}(q^m)^n$, then $E_C \left( \bigcup_i W_i \right) =
\sum_i E_C(W_i)$. Suppose $Z \df \{{\bf z} \in \mathrm{GF}(q^m)^n |
E_C(\left\{{\bf z}\right\}) > 0 \}$, i.e., $Z$ is the set of vectors
covered by at least two codewords in $C$. Note that ${\bf z} \in Z$
if and only if $|B_{\rho}({\bf z}) \cap C| \geq 2$. It can be shown
that $|Z| \leq E_C(Z) = E_C(\mathrm{GF}(q^m)^n) = |C|
V_{\rho}(q^m,n) - q^{mn}$.

Before deriving the second nontrivial lower bound, we need the
following adaptation of \cite[Lemma 8]{vanwee_jct91}. Let $C$ be a
code with length $n$ and rank covering radius $\rho$ over
$\mathrm{GF}(q^m)$. We define $A \df \{{\bf x} \in
\mathrm{GF}(q^m)^n | \dr({\bf x},C) = \rho\}$.

\begin{lemma}\label{lemma:epsilon}
For ${\bf x} \in A \backslash Z$ and $0 < \rho < n$, we have that
$E_C(B_1({\bf x})) \geq \epsilon$, where
\begin{equation}\label{eq:epsilon}
    \nonumber
    \epsilon \df \left\lceil
    \frac{(q^m-q^\rho)({n \brack 1} - {\rho \brack 1})}
    {q^{\rho}{\rho+1 \brack 1}} \right\rceil q^{\rho}{\rho+1 \brack 1}
    + (q^m-q^\rho) \left({\rho \brack 1} - {n \brack 1}\right).
\end{equation}
\end{lemma}

\begin{proof}
Since ${\bf x} \notin Z$, there is a unique ${\bf c}_0 \in C$ such
that $\dr({\bf x}, {\bf c_0}) = \rho$. By Proposition~\ref{prop:B_r}
we have $|B_{\rho}({\bf c_0}) \cap B_1({\bf x})| = 1+
(q^m-q^\rho){\rho \brack 1} + (q^\rho-1){n \brack 1}$. For any
codeword ${\bf c}_1 \in C$ satisfying $\dr({\bf x}, {\bf c}_1) =
\rho + 1$, by Proposition~\ref{prop:v-a_problem} we have
$|B_{\rho}({\bf c}_1) \cap B_1({\bf x})| = q^{\rho}{\rho+1 \brack
1}$. Finally, for all other codewords ${\bf c}_2 \in C$ at distance
$> \rho+1$ from ${\bf x}$, we have $|B_{\rho}({\bf c}_2) \cap
B_1({\bf x})| = 0$. Denoting $N \df |\{ {\bf c}_1 \in C | \dr({\bf
x},{\bf c}_1) = \rho+1 \}|$, we obtain
\begin{eqnarray*}
    E_C(B_1({\bf x})) &=& \sum_{{\bf c} \in C}|B_{\rho}({\bf c})
    \cap B_1({\bf x})| - |B_1({\bf x})|\\
    &=& (q^m-q^\rho){\rho \brack 1} + N q^{\rho}{\rho+1 \brack 1} -
    {n \brack 1}(q^m-q^\rho)\\
    &\equiv& (q^m-q^\rho)\left({\rho \brack 1} - {n \brack 1} \right)
    \mod \left( q^{\rho}{\rho+1 \brack 1} \right).
\end{eqnarray*}
The proof is completed by realizing that $(q^m-q^\rho)\left({\rho
\brack 1} - {n \brack 1} \right) < 0$, while $E_C(B_1({\bf x}))$ is
a non-negative integer.
\end{proof}

For $\rho = n-1$, Lemma~\ref{lemma:epsilon} is improved to:

\begin{corollary}\label{cor:epsilon_n-1}
For ${\bf x} \in A \backslash Z$ and $\rho = n-1$, $E_C(B_1({\bf
x})) = \phi$, where $\phi \df q^{n-1} {n \brack 1}|C| -
q^{n-1}\left(q^m + {n-1 \brack 1} \right)$.
\end{corollary}

\begin{proof}
The proof calls the same arguments as the proof above, with $N =
|C|-1$ for $\rho = n-1$.
\end{proof}

\begin{proposition}\label{prop:excess_bound}
If $\epsilon>0$, then $K_{\mbox{\tiny R}}(q^m,n,\rho) \geq
\frac{q^{mn}}{v(\rho) - \frac{\epsilon}{\delta} N_{\rho}(q^m,n)}$,
where $\delta \df v(1) - q^{\rho-1}{\rho \brack 1} -1 + 2\epsilon$.
\end{proposition}

The proof of Proposition~\ref{prop:excess_bound}, provided in
Appendix~\ref{app:prop:excess_bound}, uses the approach in the proof
of \cite[Theorem 6]{vanwee_jct91} and is based on the concept of
excess reviewed in Section~\ref{sec:covering_radius}. The lower
bounds in Propositions~\ref{prop:bound_cohen_generalised}
and~\ref{prop:excess_bound}, when applicable, are at least as tight
as the sphere covering bound. For $\rho = n-1$,
Proposition~\ref{prop:excess_bound} is refined into the following.

\begin{corollary}\label{cor:bound_excess_n-1}
Let us denote the coefficients $q^{n-1} {n \brack 1}$ and
$q^{n-1}\left(q^m + {n-1 \brack 1} \right)$ as $\alpha$ and $\beta$,
respectively. $\Kr(q^m,n,n-1)$ satisfies $a \Kr(q^m,n,n-1)^2
-b\Kr(q^m,n,n-1) + c \geq 0$, where $a \df \alpha[v(n-1) + v(n-2)]$,
$b \df v(n-1) \left\{q^{n-2}{n-1 \brack 1} - v(1) + \beta + 1
\right\} + 2\alpha q^{mn} + \beta v(n-2)$, and $c \df q^{mn} \{
2\beta + 1 + q^{n-2}{n-1 \brack 1} - v(1) \}$.
\end{corollary}

The proof of Corollary~\ref{cor:bound_excess_n-1} is given in
Appendix~\ref{app:cor:bound_excess_n-1}.

\subsection{Upper bounds for the sphere covering problem}\label{sec:upper_bounds}

From the perspective of covering, the following lemma gives a
characterization of MRD codes in terms of ELS's.

\begin{lemma}\label{lemma:C+V}
Let $\mathcal{C}$ be an $(n,k)$ linear code over $\mathrm{GF}(q^m)$
($n \leq m$). $\mathcal{C}$ is an MRD code if and only if
$\mathcal{C} \oplus \mathcal{V} = \mathrm{GF}(q^m)^n$ for all
$\mathcal{V} \in E_{n-k}(q^m,n)$.
\end{lemma}

\begin{proof}
Suppose $\mathcal{C}$ is an $(n,k,n-k+1)$ linear MRD code. It is
clear that $\mathcal{C} \cap \mathcal{V} = \left\{{\mathbf
0}\right\}$ and hence $\mathcal{C} \oplus \mathcal{V} =
\mathrm{GF}(q^m)^n$ for all $\mathcal{V} \in E_{n-k}(q^m,n)$.
Conversely, suppose $\mathcal{C} \oplus \mathcal{V} =
\mathrm{GF}(q^m)^n$ for all $\mathcal{V} \in E_{n-k}(q^m,n)$. Then
$\mathcal{C}$ does not contain any nonzero codeword of weight $\leq
n-k$, and hence its minimum distance is $n-k+1$.
\end{proof}

For $1\leq u \leq \rho$, let $\alpha_0=1, \alpha_1,
\ldots,\alpha_{m+u-1} \in \mathrm{GF}(q^{m+u})$ be a basis set of
$\mathrm{GF}(q^{m+u})$ over $\mathrm{GF}(q)$, and let $\beta_0=1,
\beta_1, \ldots, \beta_{m-1}$ be a basis of $\mathrm{GF}(q^m)$ over
$\mathrm{GF}(q)$. We define the \emph{linear} mapping $f$ between
two vector spaces $\mathrm{GF}(q^m)$ and $\mathfrak{S}_m \df
\mathfrak{S}(\alpha_0, \alpha_1, \ldots, \alpha_{m-1})$ given by
$f(\beta_i) = \alpha_i$ for $0 \leq i \leq m-1$. We remark that
$\alpha_0 = \beta_0 = 1$ implies that $f$ maps $\mathrm{GF}(q)$ to
itself. This can be generalized to $n$-dimensional vectors, by
applying $f$ componentwise. We thus define ${\bar f} :
\mathrm{GF}(q^m)^n \rightarrow \mathrm{GF}(q^{m+u})^n$ such that for
any ${\bf v} = (v_0,\ldots,v_{n-1})$, ${\bar f}({\bf v}) = (f(v_0),
\ldots, f(v_{n-1}))$. Note that ${\bar f}$ depends on $u$, but we
omit this dependence for simplicity of notation. This function
${\bar f}$ is a linear bijection from $\mathrm{GF}(q^m)^n$ to its
image $\mathfrak{S}_m^n$, and hence ${\bar f}$ preserves the rank.
$\bar{f}$ also introduces a connection between ELS's as shown below.

\begin{lemma}\label{lemma:f(V)}
For $u \geq 1$, $r\leq n$, and any $\mathcal{V} \in E_r(q^m,n)$,
${\bar f}(\mathcal{V}) \subset \mathcal{W}$, where $\mathcal{W} \in
E_r(q^{m+u},n)$. Furthermore, ${\bar f}$ induces a bijection between
$E_r(q^m,n)$ and $E_r(q^{m+u},n)$.
\end{lemma}

\begin{proof}
Let $B = \{{\bf b}_i\}$ be an elementary basis of $\mathcal{V} \in
E_r(q^m,n)$. Then, ${\bf b}_i \in \mathrm{GF}(q)^n$ and ${\bf b}_i
={\bar f}({\bf b}_i)$. Thus, $\{{\bar f}({\bf b}_i)\}$ form an
elementary basis, and hence ${\bar f}(\mathcal{V}) \subset
\mathcal{W}$, where $\mathcal{W} \in E_r(q^{m+u},n)$ with $\{{\bar
f}({\bf b}_i)\}$ as a basis. It is easy to verify that ${\bar f}$
induces a bijection between $E_r(q^m,n)$ and $E_r(q^{m+u},n)$.
\end{proof}

\begin{proposition}\label{prop:rho_MRD}
Let $\mathcal{C}$ be an $(n,n-\rho,\rho+1)$ MRD code over GF$(q^m)$
($n \leq m$) with covering radius $\rho$. For $0\leq u \leq \rho$,
the code ${\bar f}(\mathcal{C})$, where ${\bar f}$ is as defined
above, is a code of length $n$ over $\mathrm{GF}(q^{m+u})$ with
cardinality $q^{m(n-\rho)}$ and covering radius $\rho$.
\end{proposition}
\begin{proof}
The other parameters for the code are obvious, and it suffices to
establish the covering radius. Let $\mathfrak{T}_u$ be a subspace of
$\mathrm{GF}(q^{m+u})$ with dimension $u$ such that $\mathfrak{S}_m
\oplus \mathfrak{T}_u = \mathrm{GF}(q^{m+u})$. Any ${\bf u} \in
\mathrm{GF}(q^{m+u})^n$ can be expressed as ${\bf u} = {\bf v} +
{\bf w}$, where ${\bf v} \in \mathfrak{S}_m^n$ and ${\bf w} \in
\mathfrak{T}_u^n$. Hence $\rk({\bf w}) \leq u$, and ${\bf w} \in
{\mathcal W}$ for some ${\mathcal W} \in E_\rho(q^{m+u},n)$ by
Lemma~\ref{lemma:unique_ELS}. By Lemmas~\ref{lemma:C+V} and
\ref{lemma:f(V)}, we can express ${\bf v}$ as ${\bf v} = {\bar
f}({\bf c} + {\bf e}) = {\bar f}({\bf c}) + {\bar f}({\bf e})$,
where ${\bf c} \in \mathcal{C}$ and ${\bf e} \in \mathcal{V}$, such
that ${\bar f}(\mathcal{V}) \subset {\mathcal W}$. Eventually, we
have ${\bf u} = {\bar f}({\bf c}) + {\bar f}({\bf e}) + {\bf w}$,
where ${\bar f}({\bf e}) + {\bf w} \in {\mathcal W}$, and thus
$d({\bf u}, {\bar f}({\bf c})) \leq \rho$. Thus
$\bar{f}(\mathcal{C})$ has covering radius $\leq \rho$. Finally, it
is easy to verify that the covering radius of $\bar{f}(\mathcal{C})$
is exactly $\rho$.
\end{proof}

\begin{corollary}\label{cor:bound_K}
For $0 < \rho < n \leq m$, $\Kr(q^m,n,\rho) \leq q^{ \max\{m-\rho,
n\} (n-\rho)}$.
\end{corollary}
\begin{proof}
We can construct an $(n,n-\rho)$ linear MRD code $\mathcal{C}$ over
$\mathrm{GF}(q^\mu)$ with covering radius $\rho$, where $\mu =
\max\{m-\rho,n\}$. By Proposition~\ref{prop:rho_MRD}, ${\hat
f}(\mathcal{C})\subset \mathrm{GF}(q^m)^n$, where ${\hat f}$ is a
rank-preserving mapping from $\mathrm{GF}(q^\mu)^n$ to a subset of
$\mathrm{GF}(q^m)^n$ similar to ${\bar f}$ above, has covering
radius $\leq \rho$. Thus, $\Kr(q^m,n,\rho) \leq |{\hat
f}(\mathcal{C})| = |\mathcal{C}| = q^{\mu(n-\rho)}$.
\end{proof}

We use the properties of $\Kr(q^m,n,\rho)$ in
Lemma~\ref{lemma:K_n_n'} in order to obtain a tighter upper bound
when $\rho > m-n$.

\begin{proposition}\label{prop:bound_K_mixed}
Given fixed $m$, $n$, and $\rho$, for any $0 < l \leq n$ and $(n_i,
\rho_i)$ for $0\leq i \leq l-1$ so that $0 < n_i \leq n$, $0 \leq
\rho_i \leq n_i$, and $n_i + \rho_i \leq m$ for all $i$, and
$\sum_{i=0}^{l-1} n_i = n$ and $\sum_{i=0}^{l-1} \rho_i = \rho$, we
have
\begin{equation}\label{eq:bound_K_mixed}
    \Kr(q^m,n,\rho) \leq \min_{\left\{(n_i, \rho_i):
    \,0\leq i \leq l-1\right\}} \left\{ q^{m(n-\rho) - \sum_i \rho_i(n_i-\rho_i)}
    \right\}.
\end{equation}
\end{proposition}

\begin{proof}
By Lemma~\ref{lemma:K_n_n'}, we have $\Kr(q^m,n,\rho) \leq \prod_i
\Kr(q^m,n_i,\rho_i)$ for all possible sequences $\{\rho_i\}$ and
$\{n_i\}$. For all $i$, we have $\Kr(q^m,n_i,\rho_i) \leq
q^{(m-\rho_i)(n-\rho_i)}$ by Corollary~\ref{cor:bound_K}, and hence
$\Kr(q^m,n,\rho) \leq q^{\sum_i(m-\rho_i)(n_i-\rho_i)} =
q^{m(n-\rho) - \sum_i \rho_i(n_i-\rho_i)}$.
\end{proof}

It is clear that the upper bound in~(\ref{eq:bound_K_mixed}) is
tighter than the upper bound in
Proposition~\ref{prop:obvious_bounds_K}. It can also be shown that
it is tighter than the bound in Corollary~\ref{cor:bound_K}.

The following upper bound is an adaptation of \cite[Theorem
12.1.2]{cohen_book97}.

\begin{proposition}\label{prop:bound_K_12.1}
For $0 < \rho < n \leq m$, $\Kr(q^m,n,\rho) \leq \frac{1} {1-
\log_{q^{mn}}\left(q^{mn} - v(\rho) \right)} + 1.$
\end{proposition}

Our proof, given in Appendix~\ref{app:prop:bound_K_12.1}, adopts the
approach used to prove \cite[Theorem 12.1.2]{cohen_book97}. Refining
\cite[Theorem 12.2.1]{cohen_book97} for the rank metric, we obtain
the following upper bound.

\begin{proposition}\label{prop:bound_JSL}
For $0 < \rho < n \leq m$ and $a \df \min\{n,2\rho\}$,
\begin{equation}\label{eq:bound_JSL}
    \Kr(q^m,n,\rho) \leq k_{v(\rho)} \left(\frac{1}{\min\{s,j\}} - \frac{1}{v(\rho)} \right)
    + \frac{q^{mn}}{v(\rho)} H_{\min\{s,j\}},
\end{equation}
where $k_{v(\rho)} = q^{mn} - v(\rho)q^{m(n-a)}$, $j = \lceil
v(\rho) - v(\rho)^2 q^{-ma} \rceil$, $s = v(\rho) - \sum_{i =
a-\rho}^\rho q^{i(a-i)} {a \brack i}$, and $H_k \df \sum_{i=1}^k
\frac{1}{i}$ is the $k$-th harmonic number.
\end{proposition}

\begin{proof}
We denote the vectors of $\mathrm{GF}(q^m)^n$ as ${\bf v}_i$ for $i
= 0, 1, \ldots, q^{mn}-1$ and we consider a $q^{mn} \times q^{mn}$
square matrix ${\bf A}$ defined as $a_{i,j} = 1$ if $\dr({\bf v}_i,
{\bf v}_j) \leq \rho$ and $a_{i,j} = 0$ otherwise. Note that each
row and each column of ${\bf A}$ has exactly $v(\rho)$ ones. We
present an algorithm that selects $K$ columns of ${\bf A}$ with no
all-zero rows. These columns thus represent a code with cardinality
$K$ and covering radius $\rho$.

Set ${\bf A}_{v(\rho)} = {\bf A}$ and $k_{v(\rho)+1} = q^{mn}$. For
$i=v(\rho), v(\rho)-1, \ldots, 1$, repeat the following step: First,
select from ${\bf A}_i$ a maximal set of $K_i$ columns of weight $i$
with pairwise disjoint supports; Then, remove these columns and all
the $iK_i = k_{i+1}- k_i$ rows incident to one of them, and denote
the remaining $k_i \times (q^{mn}-K_{v(\rho)}-\cdots-K_i)$ matrix as
${\bf A}_{i-1}$. The set of all selected columns hence contains no
all-zero rows.

For $2\rho \leq n$, we can select an $(n,n-2\rho,2\rho+1)$ linear
MRD code $\mathcal{C}$ for the first step. Since MRD codes are
maximal, they satisfy the condition. If $2\rho > n$, $\mathcal{C}$
is chosen to be a single codeword, and $K_{v(\rho)} = 1$. Thus
$K_{v(\rho)} = q^{m(n-a)}$ and $k_{v(\rho)} = q^{mn} - v(\rho)
q^{m(n-a)}$, where $a = \min\{n,2\rho\}$.

We now establish two upper bounds on $k_i$ for $1 \leq i \leq
v(\rho)$. First, it is obvious that $k_i \leq k_v(\rho)$. Also,
every row of ${\bf A}_{i-1}$ contains exactly $v(\rho)$ ones; on the
other hand, every column of ${\bf A}_{i-1}$ contains at most $i-1$
ones. Hence for $1 \leq i \leq v(\rho)$, $v(\rho) k_i \leq
(i-1)(q^{mn} - K_{v(\rho)} - \cdots - K_i) \leq (i-1)q^{mn}$, and
thus
\begin{equation}\label{eq:ki}
    k_i \leq (i-1)\frac{q^{mn}}{v(\rho)}.
\end{equation}
Clearly $k_1 = 0$  by (\ref{eq:ki}). We have $k_{v(\rho)} \leq
(i-1)\frac{q^{mn}}{v(\rho)}$ if $i-1 \geq j \df \left\lceil
\frac{v(\rho) k_{v(\rho)}}{q^{mn}} \right\rceil.$

We now establish an upper bound on $K = \sum_{i=1}^{v(\rho)} K_i$.
For any vector ${\bf x} \in \mathrm{GF}(q^m)^n$, we have $\dr({\bf
x}, \mathcal{C}) \leq 2\rho$. This is trivial for $2\rho > n$, while
for $2\rho \leq n$ this is because MRD codes are maximal codes. For
$2\rho \leq n$, at least $q^{\rho^2} {2\rho \brack \rho}$ vectors in
$B_\rho({\bf x})$ are already covered by $\mathcal{C}$ by
Proposition~\ref{prop:v-a_problem}. For $2\rho > n$, it can be shown
that at least $\sum_{i = n-\rho}^\rho q^{i(n-i)} {n \brack i}$
vectors in $B_\rho({\bf x})$ are already covered by $\mathcal{C}$.
It follows that after the first step, the column weight is at most
$s \df v(\rho) - \sum_{i = a-\rho}^\rho q^{i(a-i)} {a \brack i}$.
Since $s \geq \max\{ 1\leq i < v(\rho) : K_i
> 0\}$, $K = K_{v(\rho)} + \sum_{i=1}^s K_i = K_{v(\rho)} + \frac{k_{v(\rho)}}{s} + \sum_{i=2}^s \frac{k_i}{i(i-1)}.$
Using the two upper bounds on $k_i$ above, we obtain $K \leq
k_{v(\rho)} \left(\frac{1}{\min\{s,j\}} - \frac{1}{v(\rho)} \right)
+ \frac{q^{mn}}{v(\rho)} H_{\min\{s,j\}}.$
\end{proof}

Following \cite{cohen_book97}, where \cite[Theorem
12.1.2]{cohen_book97} is referred to as the Johnson-Stein-Lov\'asz
theorem, we refer to the algorithm described in the proof of
Proposition~\ref{prop:bound_JSL} as the Johnson-Stein-Lov\'asz (JSL)
algorithm. The upper bound in Proposition~\ref{prop:bound_JSL} can
be loosened into the following.

\begin{corollary}\label{cor:bound_JSL_loose}
For $0 < \rho < n \leq m$, $\Kr(q^m,n,\rho) \leq
\frac{q^{mn}}{v(\rho)} \left( \ln v(\rho) + \gamma +
\frac{1}{2v(\rho) + 1/3} \right)$, where $\gamma$ is the
Euler-Mascheroni constant \cite{chen_arxiv03}.
\end{corollary}

\begin{proof}
Using~(\ref{eq:ki}), we obtain $K = \sum_{i=1}^{v(\rho)}
\frac{k_{i+1} - k_i}{i} \leq \frac{q^{mn}}{v(\rho)} +
\sum_{i=2}^{v(\rho)} \frac{k_i}{i(i-1)} \leq \frac{q^{mn}}{v(\rho)}
H_{v(\rho)}.$ The proof is concluded by $H_{v(\rho)} < \ln v(\rho) +
\gamma + \frac{1}{2v(\rho) + 1/3}$ \cite{chen_arxiv03}.
\end{proof}

\subsection{Covering properties of linear rank metric codes}
\label{sec:covering_linear}

Proposition~\ref{prop:obvious_bounds_K} yields bounds on the
dimension of a linear code with a given rank covering radius.

\begin{proposition}\label{prop:bound_k_rho}
An $(n,k)$ linear code over $\mathrm{GF}(q^m)$ with rank covering
radius $\rho$ satisfies $n- \rho - \frac{\rho(n-\rho)-\log_q K_q}{m}
< k \leq n- \rho$.
\end{proposition}

\begin{proof}
The upper bound directly follows the upper bound in
Proposition~\ref{prop:obvious_bounds_K}. We now prove the lower
bound. By the sphere covering bound, we have $q^{mk}
> \frac{q^{mn}}{v(\rho)}$. However, by
Lemma~\ref{lemma:lower_bound_Vt} we have $v(\rho) <
q^{\rho(m+n-\rho) - \log_q K_q}$ and hence $q^{mk} > q^{mn -
\rho(m+n-\rho) + \log_q K_q}$.
\end{proof}

We do not adapt the bounds in
Propositions~\ref{prop:bound_cohen_generalised}
and~\ref{prop:excess_bound} as their advantage over the lower bound
in Proposition~\ref{prop:bound_k_rho} is not significant. Next, we
show that the dimension of a linear code with a given rank covering
radius can be determined under some conditions.

\begin{proposition}\label{prop:maximal_k}
Let $\mathcal{C}$ be an $(n,k)$ linear code over $\mathrm{GF}(q^m)$
($n \leq m$) with rank covering radius $\rho$. Then $k=n-\rho$ if
$\rho \in \{0,1,n-1,n\}$ or $\rho(n-\rho) \leq m + \log_q K_q$, or
if $\mathcal{C}$ is a generalized Gabidulin code or an \ELS{}.
\end{proposition}

\begin{proof}
The cases $\rho \in \{0,n-1,n\}$ are straightforward. In all other
cases, since $k \leq n-\rho$ by Proposition~\ref{prop:bound_k_rho},
it suffices to prove that $k \geq n-\rho$. First, suppose $\rho =
1$, then $k$ satisfies $q^{mk} > \frac{q^{mn}}{v(1)}$ by the sphere
covering bound. However, $v(1) < q^{m+n} \leq q^{2m}$, and hence $k
> n-2$. Second, if $\rho(n-\rho) \leq m + \log_q K_q$, then $0<
\frac{1}{m}\left( \rho(n-\rho) - \log_q K_q \right) \leq 1$ and $k
\geq n-\rho $ by Proposition~\ref{prop:bound_k_rho}. Third, if
$\mathcal{C}$ is an $(n,k,n-k+1)$ generalized Gabidulin code with
$k<n$, then there exists an $(n,k+1,n-k)$ generalized Gabidulin code
$\mathcal{C}'$ such that $\mathcal{C} \subset \mathcal{C}'$. We have
$\rho \geq \dr(\mathcal{C}') = n-k$, as noted in
Section~\ref{sec:covering_radius}, and hence $k \geq n - \rho$. The
case $k=n$ is straightforward. Finally, if $\mathcal{C}$ is an
\ELS{} of dimension $k$, then for all ${\bf x}$ with rank $n$ and
for any ${\bf c} \in \mathcal{C}$, $\dr({\bf x}, {\bf c}) \geq
\rk({\bf x}) - \rk({\bf c}) \geq n-k$.
\end{proof}

A similar argument can be used to bound the covering radius of the
cartesian products of generalized Gabidulin codes.

\begin{corollary}\label{cor:radius_cartesian}
Let $\mathcal{G}$ be an $(n,k,d_{\mbox{\tiny R}})$ generalized
Gabidulin code $(n \leq m)$, and let $\mathcal{G}^l$ be the code
obtained by $l$ cartesian products of $\mathcal{G}$ for $l \geq 1$.
Then the rank covering radius of $\mathcal{G}^l$ satisfies
$\rho(\mathcal{G}^l) \geq \dr - 1$.
\end{corollary}

Note that when $n=m$, $\mathcal{G}^l$ is a maximal code, and hence
Corollary~\ref{cor:radius_cartesian} can be further strengthened.

\begin{corollary}\label{cor:radius_cartesian_m}
Let $\mathcal{G}$ be an $(m,k,\dr)$ generalized Gabidulin code over
$\mathrm{GF}(q^m)$, and let $\mathcal{G}^l$ be the code obtained by
$l$ cartesian products of $\mathcal{G}$. Then $\rho(\mathcal{G}^l) =
\dr-1$.
\end{corollary}

\subsection{Numerical methods}\label{sec:numerical_methods}

In addition to the above bounds, we use several different numerical
methods to obtain tighter upper bounds for relatively small values
of $m$, $n$, and $\rho$. First, the JSL algorithm described in the
proof of Proposition~\ref{prop:bound_JSL} is implemented for small
parameter values. Second, local search algorithms
\cite{cohen_book97} similar to the ones available for Hamming metric
codes are somewhat less complex than the JSL algorithm. Although the
complexity for large parameter values is prohibitive, it is
feasible. Third, we construct linear codes with good covering
properties, because linear codes have lower complexity.

We can verify if a covering radius is achievable by a given code
size by brute force verification, thereby establishing lower bounds
on $\Kr(q^m,n,\rho)$. Obviously, this is practical for only small
parameter values.

\subsection{Tables}

In Table~\ref{table:bounds}, we provide bounds on $\Kr(q^m,n,\rho)$,
for $2 \leq m \leq 7$, $2 \leq n \leq m$, and $1 \leq \rho \leq 6$.
Obviously, $\Kr(q^m,n,\rho) = 1$ when $\rho = n$. For other sets of
parameters, the tightest lower and upper bounds on $\Kr(q^m,n,\rho)$
are given, and letters associated with the numbers are used to
indicate the tightest bound. The lower case letters a--f correspond
to the lower bounds in
Propositions~\ref{prop:obvious_bounds_K},~\ref{prop:K>=3},
and~\ref{prop:bound_cohen_generalised},
Corollaries~\ref{cor:bound_K_I} and~\ref{cor:bound_cohen}, and
Proposition~\ref{prop:excess_bound} respectively. The lower case
letter g corresponds to lower bounds obtained by brute force
verification. The upper case letters A--E denote the upper bounds in
Proposition~\ref{prop:obvious_bounds_K},
Corollary~\ref{cor:bound_K}, and
Propositions~\ref{prop:bound_K_mixed},~\ref{prop:bound_K_12.1},
and~\ref{prop:bound_JSL} respectively. The upper case letters F--H
correspond to upper bounds obtained by the JSL algorithm, local
search algorithm, and explicit linear constructions respectively.

In Table~\ref{table:linear_bounds}, we provide bounds on the minimum
dimension $k$ for $q=2$, $4 \leq m \leq 8$, $4 \leq n \leq m$, and
$2 \leq \rho \leq 6$. The unmarked entries correspond to
Proposition~\ref{prop:maximal_k}. The lower case letters a and e
correspond to the lower bound in Proposition~\ref{prop:bound_k_rho}
and the adaptation of Corollary~\ref{cor:bound_cohen} to linear
codes respectively. The lower case letter h corresponds to lower
bounds obtained by brute force verification for linear codes. The
upper case letter A corresponds to the upper bound in
Proposition~\ref{prop:bound_k_rho}. The upper case letter H
corresponds to upper bounds obtained by explicit linear
constructions.

Although no analytical expression for $I(\rho,d)$ is known to us, it
can be obtained by simple counting for the bounds in
Proposition~\ref{prop:bound_cohen_generalised} or
Corollary~\ref{cor:bound_cohen}. In
Appendix~\ref{app:numerical_results}, we present the values of
$I(q^m,n,\rho,d)$ used in calculating the values of the bounds in
Proposition~\ref{prop:bound_cohen_generalised} and
Corollary~\ref{cor:bound_K_I} displayed in Table~\ref{table:bounds}.
We also present the codes, obtained by the numerical methods in
Section~\ref{sec:numerical_methods}, that achieve the tightest upper
bounds in Tables~\ref{table:bounds} and~\ref{table:linear_bounds}.

\begin{table}
\begin{center}
\begin{tabular}{|c|c|cccccc|}
    \hline
    $m$ & $n$ & $\rho=1$ & $\rho=2$ & $\rho=3$ & $\rho=4$ &
    $\rho=5$ & $\rho=6$\\
    \hline
    2 & 2 & e 3 F & 1 & & & &\\
    \hline
    3 & 2 & e 4 B & 1 & & & &\\
      & 3 & e 11-16 F & g 4 C & 1 & & &\\
    \hline
    4 & 2 & e 7-8 B & 1 & & & &\\
      & 3 & e 40-64 B & d 4-7 F & 1 & & &\\
      & 4 & f 293-722 F & e 10-48 G & b 3-7 F & 1 & &\\
    \hline
    5 & 2 & e 12-16 B & 1 & & & &\\
      & 3 & e 154-256 B & d 6-8 B & 1 & & &\\
      & 4 & e 2267-4096 B & e 33-256 C & d 4-8 C & 1 & &\\
      & 5 & e 34894-$2^{17}$ C & e 233-2881 E & a 9-32 H & b 3-8 C & 1&\\
    \hline
    6 & 2 & e 23-32 B & 1 & & & &\\
      & 3 & e 601-1024 B & d 11-16 B & 1 & &  &\\
      & 4 & e 17822-$2^{15}$ B & c 124-256 B & d 6-16 C & 1 &  &\\
      & 5 & e 550395-$2^{20}$ B & e 1770-$2^{14}$ C & f 31-256 C & a 3-16 C & 1 &\\
      & 6 & f 17318410-$2^{26}$ C & f 27065-413582 E & f 214-4211 E & f 9-181 D &
      b 3-16 C & 1\\
    \hline
    7 & 2 & e 44-64 B & 1 & & & &\\
      & 3 & e 2372-4096 B & d 20-32 B & 1 & & & \\
      & 4 & e 141231-$2^{18}$ B & f 484-1024 B & a 9-16 B & 1 & & \\
      & 5 & e 8735289-$2^{24}$ B & e 13835-$2^{15}$ B & a 111-1024 C & a 5-16 C & 1 &\\
      & 6 & e 549829402-$2^{30}$ B & f 42229-$2^{22}$ C & e 1584-$2^{15}$ C & f 29-734 E &
      a 3-16 C & 1\\
      & 7 & e 34901004402-$2^{37}$ C & f 13205450-239280759 E & e 23978-586397 E & f 203-5806 E &
      a 8-242 D & b 3-16 C\\
    \hline
\end{tabular}
\caption{Bounds on $\Kr(q^m,n,\rho)$, for $2 \leq m \leq 7$, $2 \leq
n \leq m$, and $1 \leq \rho \leq 6$.}\label{table:bounds}
\end{center}
\end{table}

\begin{table}[!htp]
\begin{center}
\begin{tabular}{|c|c|ccccc|}
    \hline
    $m$ & $n$ & $\rho=2$ & $\rho=3$ & $\rho=4$ & $\rho=5$ &
    $\rho=6$\\
    \hline
    4 & 4 & h 2 A & 1 & 0 & & \\
    \hline
    5 & 4 & e 2 A & 1 & 0 & & \\
      & 5 & a 2-3 A & a 1 H & 1 & 0 & \\
    \hline
    6 & 4 & 2 & 1 & 0 & & \\
      & 5 & a 2-3 A & a 1-2 A & 1 & 0 & \\
      & 6 & a 3-4 A & a 2-3 A & a 1-2 A & 1 & 0\\
    \hline
    7 & 4 & 2 & 1 & 0 & & \\
      & 5 & a 2-3 A & a 1-2 A & 1 & 0 & \\
      & 6 & a 3-4 A & a 2-3 A & a 1-2 A & 1 & 0\\
      & 7 & a 4-5 A & a 3-4 A & a 2-3 A & a 1-2 A & 1\\
    \hline
    8 & 4 & 2 & 1 & 0 & & \\
      & 5 & 3 & 2 & 1 & 0 & \\
      & 6 & a 3-4 A & a 2-3 A & a 1-2 A & 1 & 0\\
      & 7 & a 4-5 A & a 3-4 A & a 2-3 A & a 1-2 A & 1\\
      & 8 & a 5-6 A & a 3-5 A & a 2-4 A & a 1-3 A & a 1-2 A\\
    \hline
\end{tabular}
\caption{Bounds on $k$ for $q=2$, $4 \leq m \leq 8$, $4 \leq n \leq
m$, and $2 \leq \rho \leq 6$.}\label{table:linear_bounds}
\end{center}
\end{table}

\subsection{Asymptotic covering properties}\label{sec:asym_covering}
Table~\ref{table:bounds} provides solutions to the sphere covering
problem for only small values of $m$, $n$, and $\rho$. Next, we
study the asymptotic covering properties when both block length and
minimum rank distance go to infinity. As in
Section~\ref{sec:packing}, we consider the case where $\lim_{n
\rightarrow \infty}\frac{n}{m}=b$, where $b$ is a constant. In other
words, these asymptotic covering properties provide insights on the
covering properties of long rank metric codes over large fields.

The asymptotic form of the bounds in
Lemma~\ref{lemma:lower_bound_Vt} are given in the lemma below.
\begin{lemma}\label{lemma:v(delta)}
For $0 \leq \delta \leq \min\{1,b^{-1}\}$, $\lim_{n \rightarrow
\infty} \left[ \log_{q^{mn}} V_{\lfloor\delta
n\rfloor}(q^m,n)\right] = \delta(1+b-b\delta)$.
\end{lemma}

\begin{proof}
By Lemma~\ref{lemma:lower_bound_Vt}, we have $q^{\dr(m+n-\dr)} \leq
v(\dr) < K_q^{-1} q^{\dr(m+n-\dr)}$. Taking the logarithm, this
becomes $\delta(1+b-b\delta) \leq \log_{q^{mn}} v(\lfloor\delta
n\rfloor) < \delta (1+b-b\delta) - \frac{\log_q K_q}{mn}$. The proof
is concluded by taking the limit when $n$ tends to infinity.
\end{proof}

Define $r \df \frac{\rho}{n}$ and $k(r) = \lim_{n \rightarrow
\infty} \inf \left[ \log_{q^{mn}}\Kr(q^m,n,\rho) \right]$. The
bounds in Proposition~\ref{prop:obvious_bounds_K} and
Corollary~\ref{cor:bound_JSL_loose} together solve the asymptotic
sphere covering problem.

\begin{theorem}\label{th:asymptotic_covering}
For all $b$ and $r$, $k(r) = (1-r)(1-br)$.
\end{theorem}

\begin{proof}
By Lemma~\ref{lemma:v(delta)} the sphere covering bound
asymptotically becomes $k(r) \geq (1-r) (1-br)$. Also, by
Corollary~\ref{cor:bound_JSL_loose}, $\Kr(q^m,n,\rho) \leq
\frac{q^{mn}}{v(\rho)}
    \left[ 1 + \ln v(\rho) \right] \leq \frac{q^{mn}}{v(\rho)}
    \left[ 1 + mn \ln q \right]$ and hence $\log_{q^{mn}} \Kr(q^m,n,\rho) \leq \log_{q^{mn}}
    \frac{q^{mn}}{v(\rho)} + O((mn)^{-1} \ln mn)$.
By Lemma~\ref{lemma:v(delta)}, this asymptotically becomes $k(r)
\leq (1-r)(1-br)$. Note that although we assume $n \leq m$ above for
convenience, both bounds in Proposition~\ref{prop:obvious_bounds_K}
and Corollary~\ref{cor:bound_JSL_loose} hold for any values of $m$
and $n$.
\end{proof}

\appendix

\subsection{Proof of Proposition~\ref{prop:excess_bound}}\label{app:prop:excess_bound}
We first establish a key lemma.
\begin{lemma}\label{lemma:A_cap_B}
If ${\bf z} \in Z$ and $0 < \rho < n$, then $|A \cap B_1({\bf z})|
\leq v(1) - q^{\rho-1}{\rho \brack 1}.$
\end{lemma}

\begin{proof}
By definition of $\rho$, there exists ${\bf c} \in C$ such that
$\dr({\bf z},{\bf c}) \leq \rho$. By
Proposition~\ref{prop:inter_2_balls}, $|B_1({\bf z}) \cap
B_{\rho-1}({\bf c})|$ gets its minimal value for $\dr({\bf z},{\bf
c}) = \rho$, which is $q^{\rho-1}{\rho \brack 1}$ by
Proposition~\ref{prop:v-a_problem}. A vector at distance $\leq \rho
- 1$ from any codeword does not belong to $A$. Therefore, $B_1({\bf
z}) \cap B_{\rho-1}({\bf c}) \subseteq B_1({\bf z})\backslash A$,
and hence $|A \cap B_1({\bf z})| = |B_1({\bf z})| - |B_1({\bf z})
\backslash A| \leq v(1) - |B_1({\bf z}) \cap B_{\rho-1}({\bf c})|$.
\end{proof}

We now give a proof of Proposition~\ref{prop:excess_bound}.

\begin{proof}For a code $C$ with covering radius $\rho$ and $\epsilon \geq
1$,
\begin{eqnarray}
    \label{eq:gamma}
    \gamma & \df & \epsilon \left[ q^{mn} - |C|v(\rho-1) \right]
    - (\epsilon-1) \left[|C|v(\rho) - q^{mn}\right]\\
    \label{eq:vanwee1}
    & \leq & \epsilon |A| - (\epsilon-1)|Z|\\
    \nonumber
    & \leq & \epsilon |A| - (\epsilon-1) |A \cap Z|
    = \epsilon |A \backslash Z| + |A \cap Z|,
\end{eqnarray}
where~(\ref{eq:vanwee1}) follows from $|Z| \leq |C|v(\rho) -
q^{mn}$, given in Section~\ref{sec:covering_radius}.

\begin{eqnarray}
    \label{eq:vanwee2}
    \gamma & \leq & \sum_{{\bf a} \in A \backslash Z} E_C(B_1({\bf a})) + \sum_{{\bf a} \in A \cap Z}
    E_C(B_1({\bf a}))\\
    \nonumber
    & = & \sum_{{\bf a} \in A} E_C(B_1({\bf a})),
\end{eqnarray}
where~(\ref{eq:vanwee2}) follows from Lemma~\ref{lemma:epsilon} and
$|A \cap Z| \leq E_C(A \cap Z)$.

\begin{eqnarray}
    \label{eq:vanwee3}
    \gamma & \leq & \sum_{{\bf a} \in A} \sum_{{\bf x} \in B_1({\bf a}) \cap
    Z} E_C(\{ {\bf x} \})\\
    \nonumber
    & = & \sum_{{\bf x} \in Z} \sum_{{\bf a} \in B_1({\bf x}) \cap
    A} E_C(\{ {\bf x} \})
    = \sum_{{\bf x} \in Z} |A \cap B_1({\bf x})| E_C(\{ {\bf x}
    \}),
\end{eqnarray}
where~(\ref{eq:vanwee3}) follows the fact that the second summation
is over disjoint sets $\{ {\bf x} \}$. By Lemma~\ref{lemma:A_cap_B},
we obtain
\begin{eqnarray}
    \nonumber
    \gamma & \leq & \sum_{{\bf x} \in Z} \left(v(1) - q^{\rho-1}{\rho \brack 1}\right) E_C(\{ {\bf x}
    \})\\
    \nonumber
    & = & \left(v(1) - q^{\rho-1}{\rho \brack 1}\right) E_C(Z)\\
    \label{eq:vanwee4}
    & = & \left(v(1) - q^{\rho-1}{\rho \brack 1}\right)(|C|v(\rho) - q^{mn}).
\end{eqnarray}
Combining~(\ref{eq:vanwee4}) and~(\ref{eq:gamma}), we obtain the
bound in Proposition~\ref{prop:excess_bound}.
\end{proof}

\subsection{Proof of Corollary~\ref{cor:bound_excess_n-1}}\label{app:cor:bound_excess_n-1}

For $\rho = n-1$, (\ref{eq:vanwee4}) becomes $\phi \left[ q^{mn} -
|C|v(n-2) \right] - (\phi-1) \left[|C|v(n-1) - q^{mn}\right]\leq
\big(v(1) - q^{n-2}{n-1 \brack 1}\big) (|C|v(n-1) - q^{mn}).$
Substituting $\phi = \alpha |C| - \beta$ and rearranging, we obtain
the quadratic inequality in Corollary~\ref{cor:bound_excess_n-1}.

\subsection{Proof of Proposition~\ref{prop:bound_K_12.1}}
\label{app:prop:bound_K_12.1} Given a radius $\rho$ and a code $C$,
denote the set of vectors in $\mathrm{GF}(q^m)^n$ at distance $>
\rho$ from $C$ as $P_\rho(C)$. To simplify notations, $Q \df q^{mn}$
and $p_\rho(C) \df Q^{-1} |P_\rho(C)|$. Let us denote the set of all
codes over $\mathrm{GF}(q^m)$ of length $n$ and cardinality $K$ as
$S_K$. Clearly $|S_K| = {Q \choose K}$. The average value of
$p_\rho(C)$ for all codes $C \in S_K$ is given by
\begin{eqnarray}
    \nonumber
    \frac{1}{|S_K|} \sum_{C \in S_K} p_\rho(C) &=& \frac{1}{|S_K|} Q^{-1} \sum_{C \in S_K} |P_\rho(C)|
    = \frac{1}{|S_K|} Q^{-1} \sum_{C \in S_K} \sum_{{\bf x} \in F | \dr({\bf x}, C) > \rho}
    1\\
    \nonumber
    &=& \frac{1}{|S_K|} Q^{-1} \sum_{{\bf x} \in F} \sum_{C \in S_K| \dr({\bf x}, C) >
    \rho} 1\\
    \label{eq:Q_choose_K}
    &=& \frac{1}{|S_K|} Q^{-1} \sum_{{\bf x} \in F} {Q - v(\rho) \choose K}\\
    \nonumber
    &=& {Q - v(\rho) \choose K} \left/ {Q \choose K} \right.
\end{eqnarray}
Eq.~(\ref{eq:Q_choose_K}) comes from the fact that there are ${Q -
v(\rho) \choose K}$ codes with cardinality $K$ that do not cover
${\bf x}$. For all $K$, there exists a code $C' \in S_K$ for which
$p_\rho(C')$ is no more than the average, that is:
\begin{eqnarray}
    \nonumber
    p_\rho(C') &\leq& {Q \choose K}^{-1} {Q - v(\rho) \choose K}
    \leq \left( 1-Q^{-1}v(\rho)\right)^K.
\end{eqnarray}

Let us choose $K = \left\lfloor  -\frac{1}{\log_Q
\left(1-Q^{-1}v(\rho) \right)} \right\rfloor + 1$ so that $K \log_Q
\left(1 - Q^{-1} v(\rho) \right) < -1$ and hence $p_\rho(C') =
\left(1- Q^{-1} v(\rho) \right)^K < Q^{-1}$. It follows that
$|P_\rho(C')| < 1$, and $C'$ has covering radius at most $\rho$.

\subsection{Numerical results}\label{app:numerical_results}

The values of $I(q^m,n,\rho,d)$, used in calculating the bounds in
Proposition~\ref{prop:bound_cohen_generalised} or
Corollary~\ref{cor:bound_K_I}, obtained by counting are
$I(2^4,3,2,3) = 560$, $I(2^5,3,2,3) = 1232$, $I(2^5,4,3,4) = 31040$,
$I(2^6,3,2,3) = 2576$, $I(2^6,4,2,3) = 2912$, $I(2^6,4,3,4) =
756800$, and $I(2^7,3,2,3) = 5264$.

We now present the codes, obtained by computer search, that achieve
the tightest upper bounds in Tables~\ref{table:bounds}
and~\ref{table:linear_bounds}. The finite fields use the default
generator polynomials from MATLAB \cite{gf_matlab}. First, the
linear code used to show that $\Kr(2^5,5,3) \leq 32$ has a generator
matrix given by ${\bf G} = (1,\alpha,\alpha^2,0,0)$, where $\alpha$
is a primitive element of $\mathrm{GF}(2^5)$. We use the {\em
skip-vector} form \cite{brouwer_it90} to represent the other codes
obtained by computer search. The skip-vector form of a code $C =
\{{\bf c}_i\}_{i=0}^{K-1}$ over $\mathrm{GF}(q^m)^n$ can be obtained
as follows. First, each codeword ${\bf c}_i \in \mathrm{GF}(q^m)^n$
is represented by an integer $x_i$ in $[0, q^{mn}-1]$ according to
the lexicographical order. Second, the integers $x_i$ are sorted in
ascending order; the resulting integers are denoted as $x'_i$.
Third, calculate $y_i$ defined as $y_0 = x'_0$ and $y_i = x'_i -
x'_{i-1} - 1$ for $1 \leq i \leq K-1$. Fourth, if $y_i = y_{i+1} =
\ldots = y_{i+k-1}$, then we write $y_i^k$.

Below are the codes obtained by the JSL algorithm.

\noindent${\bf \Kr(2^2,2,1) = 3}\qquad$ 0$^3$

\noindent${\bf \Kr(2^3,3,1) \leq 16}\qquad$ 5    25    66    21 51 9
20     5    85    21     2    25 49     9    84     5

\noindent${\bf \Kr(2^4,3,2) \leq 7}\qquad$ 135   689    34   420 477
522 759

\noindent${\bf \Kr(2^4,4,1) \leq 722}\qquad$
     0    57   308   349    86   125    27   192    18    38    36    95    86    64   157    67    98     7   301    21
   131    42    39    60   149    97    40   116    20    85    11    90    15    56    19   167     9   137    10     7
    75    21    51    18   110    12    82    27    38    15   143     2    24   120     5    39    77   223    14    52
    27    12   179    42    86    88   100     3   130    34    66    35     5    30   171   210   137    34    29   149
    59    69    97    26   105    93   286    61    14    30   136    62     0   148    97   132   182   184     3    69
    43    31    74     5   190    85    92    85    91   146    58    75    20     4   234   292    56    10    40    56
    86    37    85   111    80    32   103    69    82    34   106   187    42    44    47   242   220    43    10   144
    27    50    97   118    60    94    61   297    36    12   222    19    16    88    72   170    19    14   197    20
   120   136    54    20    59    47    86    49    37    70   216   164$^2$      92    53    77    83    70   225    73
    38   119    33   224    34   316     1    51    36    74    33    19   128    60    52   160    31    62   135    50
   135   282    19    38   140    80    88    55    65    50    46    22    16   320    15   110    58   183   106     0
    30   170   128    82     2   152   189    60    62    61   180    30    74    22    15   201    16   184    44   206
    59    93    16   148    12    94    33   102    40    68    52    12   114    32   216    45   134    31   140    29
   324    87    97   206    14    26    42     4    22    48    89    60    85    29    14   203    37     7   300   165
   128    58   224    80    95     3    22    98    90     4   337     6    25   121    64    54    84    13   109    87
    30    49    32    56    26   116    40   126   109    47    27   100    68    14    98    60   167    33    90   224
     6   229   262    89    48    89    63   157   107    21    28   445     2    13    26   132     7    36     3    81
    11    50    43    35   127    89     7   180    26    22    89    82    18   113   230    49   278   197   323    24
    93   230   144    99    15     8   255    27     9    19    79    80    56   175   107    40    62   105    20   115
     9    41    95    72    97   109   250    51   166    47    65    94     7   166   133   108   148    56    76   201
    69    98   133    33    46    13    36   176    12    44    20    23    90    96    98   191    56    90   162    66
    39    44   107   198     0    90   124   353   354   242    21   170   161    35   211     9    14     5   155    13
    20     4   120    24    89    36    73   139    98   114   128    30    64    33    67   132    15   102   105    22
    48   161    36    35    53    19   102   150     4    30    54    18   119    14    19     0    60    84     2    50
    62    40    95    13    33   140    38    28   116    60     0   167    44   104   244   366    93    87     9   282
   157   158   248    19     7   123   218$^2$     130   236   178     0    13    12    46    97    67    30    98    25
    26    49   111     0    40    36   197     2    58    67    18    98   155    21    34     9    93   101    61     8
   111    71    68   112   232    69   403     9   148    40   237   248    99    93   230    53   171    49    89   131
    13   110    27   157   107    58    19    16    19    92   110   366    68    81   198   212    73    57   193   158
    33   123   129    52    85    23   181    48    85   150   200    73    74    41    36   183    79    72   278   145
   240    26    27   144    49   212    99    82   173    93     0   221   118    33   108    39    11$^2$     122    20
     4    12   136   177    45    39    51     6   150    30    50    10   228     4   146    77     0    14    78   117
    88   141    39   260   358     1    97   170    39   248   116    30   118    15    11    49   271    83     8   118
    32    54    96    21    67    71   234    97   229   106    59   166    19    35   152    42    56   317    11   184
    90     2    60    65    15   272   231   121    56    53    11    93   250   272    38    26    88     6   110    59
   158    14   109    29   110   113    58   206    87    46   162    99    13    22    59   220   146   161   152    73
    69   162

Below is the code obtained by a local search algorithm.

\noindent${\bf \Kr(2^4,4,2) \leq 48} \qquad$
        1493        1124         265         285        1030        2524        1366         493        6079         968
        2145         848         312         473        1307         712        1088        2274        1380        1114
        1028         567         422        1462         699         203         180        4669         146         978
        3933        1810        2083         345         354         659        1054        2314        1443        2660
        2675        1512         756        1229          95        2144        1624        1148

%
%
%
%
%
%

\bibliographystyle{IEEETran}
\bibliography{gpt}

\end{document}